
%
\catcode`@=11 
%
%
%

\font\fourteenrm=cmr10 scaled\magstep2
\font\twelverm=cmr10 scaled\magstep1
\font\ninerm=cmr9            \font\sixrm=cmr6

\font\fourteenbf=cmbx10 scaled\magstep2
\font\twelvebf=cmbx10 scaled\magstep1
\font\ninebf=cmbx9            \font\sixbf=cmbx6
\font\seventeeni=cmmi10 scaled\magstep3     \skewchar\seventeeni='177
\font\fourteeni=cmmi10 scaled\magstep2      \skewchar\fourteeni='177
\font\twelvei=cmmi10 scaled\magstep1        \skewchar\twelvei='177
\font\ninei=cmmi9                           \skewchar\ninei='177
\font\sixi=cmmi6                            \skewchar\sixi='177
\font\seventeensy=cmsy10 scaled\magstep3    \skewchar\seventeensy='60
\font\fourteensy=cmsy10 scaled\magstep2     \skewchar\fourteensy='60
\font\twelvesy=cmsy10 scaled\magstep1       \skewchar\twelvesy='60
\font\ninesy=cmsy9                          \skewchar\ninesy='60
\font\sixsy=cmsy6                           \skewchar\sixsy='60

\font\fourteenex=cmex10 scaled\magstep2
\font\twelveex=cmex10 scaled\magstep1

\font\fourteensl=cmsl10 scaled\magstep2
\font\twelvesl=cmsl10 scaled\magstep1
\font\ninesl=cmsl9

\font\fourteenit=cmti10 scaled\magstep2
\font\twelveit=cmti10 scaled\magstep1
\font\twelvett=cmtt10 scaled\magstep1
\font\twelvecp=cmcsc10 scaled\magstep1
\font\tencp=cmcsc10
\newfam\cpfam
%
%
\newcount\f@ntkey            \f@ntkey=0
\def\samef@nt{\relax \ifcase\f@ntkey \rm \or\oldstyle \or\or
         \or\it \or\sl \or\bf \or\tt \or\caps \fi }
\def\fourteenpoint{\relax
    \textfont0=\fourteenrm          \scriptfont0=\tenrm
    \scriptscriptfont0=\sevenrm
     \def\rm{\fam0 \fourteenrm \f@ntkey=0 }\relax
    \textfont1=\fourteeni           \scriptfont1=\teni
    \scriptscriptfont1=\seveni
     \def\oldstyle{\fam1 \fourteeni\f@ntkey=1 }\relax
    \textfont2=\fourteensy          \scriptfont2=\tensy
    \scriptscriptfont2=\sevensy
    \textfont3=\fourteenex     \scriptfont3=\fourteenex
    \scriptscriptfont3=\fourteenex
    \def\it{\fam\itfam \fourteenit\f@ntkey=4 }\textfont\itfam=\fourteenit
    \def\sl{\fam\slfam \fourteensl\f@ntkey=5 }\textfont\slfam=\fourteensl
    \scriptfont\slfam=\tensl
    \def\bf{\fam\bffam \fourteenbf\f@ntkey=6 }\textfont\bffam=\fourteenbf
    \scriptfont\bffam=\tenbf     \scriptscriptfont\bffam=\sevenbf
    \def\tt{\fam\ttfam \twelvett \f@ntkey=7 }\textfont\ttfam=\twelvett
    \h@big=11.9\p@{} \h@Big=16.1\p@{} \h@bigg=20.3\p@{} \h@Bigg=24.5\p@{}
    \def\caps{\fam\cpfam \twelvecp \f@ntkey=8 }\textfont\cpfam=\twelvecp
    \setbox\strutbox=\hbox{\vrule height 12pt depth 5pt width\z@}
    \samef@nt}
\def\twelvepoint{\relax
    \textfont0=\twelverm          \scriptfont0=\ninerm
    \scriptscriptfont0=\sixrm
     \def\rm{\fam0 \twelverm \f@ntkey=0 }\relax
    \textfont1=\twelvei           \scriptfont1=\ninei
    \scriptscriptfont1=\sixi
     \def\oldstyle{\fam1 \twelvei\f@ntkey=1 }\relax
    \textfont2=\twelvesy          \scriptfont2=\ninesy
    \scriptscriptfont2=\sixsy
    \textfont3=\twelveex          \scriptfont3=\twelveex
    \scriptscriptfont3=\twelveex
    \def\it{\fam\itfam \twelveit \f@ntkey=4 }\textfont\itfam=\twelveit
    \def\sl{\fam\slfam \twelvesl \f@ntkey=5 }\textfont\slfam=\twelvesl
    \scriptfont\slfam=\ninesl
    \def\bf{\fam\bffam \twelvebf \f@ntkey=6 }\textfont\bffam=\twelvebf
    \scriptfont\bffam=\ninebf     \scriptscriptfont\bffam=\sixbf
    \def\tt{\fam\ttfam \twelvett \f@ntkey=7 }\textfont\ttfam=\twelvett
    \h@big=10.2\p@{}
    \h@Big=13.8\p@{}
    \h@bigg=17.4\p@{}
    \h@Bigg=21.0\p@{}
    \def\caps{\fam\cpfam \twelvecp \f@ntkey=8 }\textfont\cpfam=\twelvecp
    \setbox\strutbox=\hbox{\vrule height 10pt depth 4pt width\z@}
    \samef@nt}
\def\tenpoint{\relax
    \textfont0=\tenrm          \scriptfont0=\sevenrm
    \scriptscriptfont0=\fiverm
    \def\rm{\fam0 \tenrm \f@ntkey=0 }\relax
    \textfont1=\teni           \scriptfont1=\seveni
    \scriptscriptfont1=\fivei
    \def\oldstyle{\fam1 \teni \f@ntkey=1 }\relax
    \textfont2=\tensy          \scriptfont2=\sevensy
    \scriptscriptfont2=\fivesy
    \textfont3=\tenex          \scriptfont3=\tenex
    \scriptscriptfont3=\tenex
    \def\it{\fam\itfam \tenit \f@ntkey=4 }\textfont\itfam=\tenit
    \def\sl{\fam\slfam \tensl \f@ntkey=5 }\textfont\slfam=\tensl
    \def\bf{\fam\bffam \tenbf \f@ntkey=6 }\textfont\bffam=\tenbf
    \scriptfont\bffam=\sevenbf     \scriptscriptfont\bffam=\fivebf
    \def\tt{\fam\ttfam \tentt \f@ntkey=7 }\textfont\ttfam=\tentt
    \def\caps{\fam\cpfam \tencp \f@ntkey=8 }\textfont\cpfam=\tencp
    \setbox\strutbox=\hbox{\vrule height 8.5pt depth 3.5pt width\z@}
    \samef@nt}
%
%
%
%
\newdimen\h@big  \h@big=8.5\p@
\newdimen\h@Big  \h@Big=11.5\p@
\newdimen\h@bigg  \h@bigg=14.5\p@
\newdimen\h@Bigg  \h@Bigg=17.5\p@
\def\big#1{{\hbox{$\left#1\vbox to\h@big{}\right.\n@space$}}}
\def\Big#1{{\hbox{$\left#1\vbox to\h@Big{}\right.\n@space$}}}
\def\bigg#1{{\hbox{$\left#1\vbox to\h@bigg{}\right.\n@space$}}}
\def\Bigg#1{{\hbox{$\left#1\vbox to\h@Bigg{}\right.\n@space$}}}
%
%
%
\normalbaselineskip = 20pt plus 0.2pt minus 0.1pt
\normallineskip = 1.5pt plus 0.1pt minus 0.1pt
\normallineskiplimit = 1.5pt
\newskip\normaldisplayskip
\normaldisplayskip = 20pt plus 5pt minus 10pt
\newskip\normaldispshortskip
\normaldispshortskip = 6pt plus 5pt
\newskip\normalparskip
\normalparskip = 6pt plus 2pt minus 1pt
\newskip\skipregister
\skipregister = 5pt plus 2pt minus 1.5pt
\newif\ifsingl@    \newif\ifdoubl@
\newif\iftwelv@    \twelv@true
\def\singlespace{\singl@true\doubl@false\spaces@t}
\def\doublespace{\singl@false\doubl@true\spaces@t}
\def\normalspace{\singl@false\doubl@false\spaces@t}
\def\Tenpoint{\tenpoint\twelv@false\spaces@t}
\def\Twelvepoint{\twelvepoint\twelv@true\spaces@t}
\def\spaces@t{\relax%
 \iftwelv@ \ifsingl@\subspaces@t3:4;\else\subspaces@t29:31;\fi%
 \else \ifsingl@\subspaces@t3:5;\else\subspaces@t4:5;\fi \fi%
 \ifdoubl@ \multiply\baselineskip by 5%
 \divide\baselineskip by 4 \fi \unskip}
\def\subspaces@t#1:#2;{
      \baselineskip = \normalbaselineskip
      \multiply\baselineskip by #1 \divide\baselineskip by #2
      \lineskip = \normallineskip
      \multiply\lineskip by #1 \divide\lineskip by #2
      \lineskiplimit = \normallineskiplimit
      \multiply\lineskiplimit by #1 \divide\lineskiplimit by #2
      \parskip = \normalparskip
      \multiply\parskip by #1 \divide\parskip by #2
      \abovedisplayskip = \normaldisplayskip
      \multiply\abovedisplayskip by #1 \divide\abovedisplayskip by #2
      \belowdisplayskip = \abovedisplayskip
      \abovedisplayshortskip = \normaldispshortskip
      \multiply\abovedisplayshortskip by #1
        \divide\abovedisplayshortskip by #2
      \belowdisplayshortskip = \abovedisplayshortskip
      \advance\belowdisplayshortskip by \belowdisplayskip
      \divide\belowdisplayshortskip by 2
      \smallskipamount = \skipregister
      \multiply\smallskipamount by #1 \divide\smallskipamount by #2
      \medskipamount = \smallskipamount \multiply\medskipamount by 2
      \bigskipamount = \smallskipamount \multiply\bigskipamount by 4 }
\def\normalbaselines{ \baselineskip=\normalbaselineskip
   \lineskip=\normallineskip \lineskiplimit=\normallineskip
   \iftwelv@\else \multiply\baselineskip by 4 \divide\baselineskip by 5
     \multiply\lineskiplimit by 4 \divide\lineskiplimit by 5
     \multiply\lineskip by 4 \divide\lineskip by 5 \fi }
\Twelvepoint  
\interlinepenalty=50
\interfootnotelinepenalty=5000
\predisplaypenalty=9000
\postdisplaypenalty=500
\hfuzz=1pt
\vfuzz=0.2pt
%
%
%
\def\pagecontents{
   \ifvoid\topins\else\unvbox\topins\vskip\skip\topins\fi
   \dimen@ = \dp255 \unvbox255
   \ifvoid\footins\else\vskip\skip\footins\footrule\unvbox\footins\fi
   \ifr@ggedbottom \kern-\dimen@ \vfil \fi }
\def\makeheadline{\vbox to 0pt{ \skip@=\topskip
      \advance\skip@ by -12pt \advance\skip@ by -2\normalbaselineskip
      \vskip\skip@ \line{\vbox to 12pt{}\the\headline} \vss
      }\nointerlineskip}
\def\makefootline{\baselineskip = 1.5\normalbaselineskip
                 \line{\the\footline}}
\newif\iffrontpage
\newif\ifletterstyle
\newif\ifp@genum
\def\nopagenumbers{\p@genumfalse}
\def\pagenumbers{\p@genumtrue}
\pagenumbers
\newtoks\paperheadline
\newtoks\letterheadline
\newtoks\letterfrontheadline
\newtoks\lettermainheadline
\newtoks\paperfootline
\newtoks\letterfootline
\newtoks\date
\footline={\ifletterstyle\the\letterfootline\else\the\paperfootline\fi}
\paperfootline={\hss\iffrontpage\else\ifp@genum\tenrm\folio\hss\fi\fi}
\letterfootline={\hfil}
\headline={\ifletterstyle\the\letterheadline\else\the\paperheadline\fi}
\paperheadline={\hfil}
\letterheadline{\iffrontpage\the\letterfrontheadline
     \else\the\lettermainheadline\fi}
\lettermainheadline={\rm\ifp@genum page \ \folio\fi\hfil\the\date}
\def\monthname{\relax\ifcase\month 0/\or January\or February\or
   March\or April\or May\or June\or July\or August\or September\or
   October\or November\or December\else\number\month/\fi}
\date={\monthname\ \number\day, \number\year}
\countdef\pagenumber=1  \pagenumber=1
\def\advancepageno{\global\advance\pageno by 1
   \ifnum\pagenumber<0 \global\advance\pagenumber by -1
    \else\global\advance\pagenumber by 1 \fi \global\frontpagefalse }
\def\folio{\ifnum\pagenumber<0 \romannumeral-\pagenumber
           \else \number\pagenumber \fi }
\def\footrule{\dimen@=\prevdepth\nointerlineskip
   \vbox to 0pt{\vskip -0.25\baselineskip \hrule width 0.35\hsize \vss}
   \prevdepth=\dimen@ }
\newtoks\foottokens
\foottokens={\Tenpoint\singlespace}
\newdimen\footindent
\footindent=24pt
\def\vfootnote#1{\insert\footins\bgroup  \the\foottokens
   \interlinepenalty=\interfootnotelinepenalty \floatingpenalty=20000
   \splittopskip=\ht\strutbox \boxmaxdepth=\dp\strutbox
   \leftskip=\footindent \rightskip=\z@skip
   \parindent=0.5\footindent \parfillskip=0pt plus 1fil
   \spaceskip=\z@skip \xspaceskip=\z@skip
   \Textindent{$ #1 $}\footstrut\futurelet\next\fo@t}
\def\Textindent#1{\noindent\llap{#1\enspace}\ignorespaces}
\def\footnote#1{\attach{#1}\vfootnote{#1}}

\let\footsymbol=\star
\newcount\lastf@@t           \lastf@@t=-1
\newcount\footsymbolcount    \footsymbolcount=0
\newif\ifPhysRev
\def\footsymbolgen{\relax \ifPhysRev \iffrontpage \NPsymbolgen\else
      \PRsymbolgen\fi \else \NPsymbolgen\fi
   \global\lastf@@t=\pageno \footsymbol }
\def\NPsymbolgen{\ifnum\footsymbolcount<0 \global\footsymbolcount=0\fi
   {\iffrontpage \else \advance\lastf@@t by 1 \fi
    \ifnum\lastf@@t<\pageno \global\footsymbolcount=0
     \else \global\advance\footsymbolcount by 1 \fi }
   \ifcase\footsymbolcount \fd@f\star\or \fd@f\dagger\or \fd@f\ast\or
    \fd@f\ddagger\or \fd@f\natural\or \fd@f\diamond\or \fd@f\bullet\or
    \fd@f\nabla\else \fd@f\dagger\global\footsymbolcount=0 \fi }
\def\fd@f#1{\xdef\footsymbol{#1}}
\def\PRsymbolgen{\ifnum\footsymbolcount>0 \global\footsymbolcount=0\fi
      \global\advance\footsymbolcount by -1
      \xdef\footsymbol{\sharp\number-\footsymbolcount} }
\def\space@ver#1{\let\@sf=\empty \ifmmode #1\else \ifhmode
   \edef\@sf{\spacefactor=\the\spacefactor}\unskip${}#1$\relax\fi\fi}
\def\attach#1{\space@ver{\strut^{\mkern 2mu #1} }\@sf\ }
\def\atttach#1{\space@ver{\strut{\mkern 2mu #1} }\@sf\ }
%
%
%
\newcount\chapternumber      \chapternumber=0
\newcount\sectionnumber      \sectionnumber=0
\newcount\equanumber         \equanumber=0
\let\chapterlabel=0
\newtoks\chapterstyle        \chapterstyle={\Number}
\newskip\chapterskip         \chapterskip=\bigskipamount
\newskip\sectionskip         \sectionskip=\medskipamount
\newskip\headskip            \headskip=8pt plus 3pt minus 3pt
\newdimen\chapterminspace    \chapterminspace=15pc
\newdimen\sectionminspace    \sectionminspace=10pc
\newdimen\referenceminspace  \referenceminspace=25pc
\def\chapterreset{\global\advance\chapternumber by 1
   \ifnum\the\equanumber<0 \else\global\equanumber=0\fi
   \sectionnumber=0 \makel@bel}
\def\makel@bel{\xdef\chapterlabel{%
\the\chapterstyle{\the\chapternumber}.}}
\def\sectionlabel{\number\sectionnumber \quad }
\def\alphabetic#1{\count255='140 \advance\count255 by #1\char\count255}
\def\Alphabetic#1{\count255='100 \advance\count255 by #1\char\count255}
\def\Roman#1{\uppercase\expandafter{\romannumeral #1}}
\def\roman#1{\romannumeral #1}
\def\Number#1{\number #1}
\def\unnumberedchapters{\let\makel@bel=\relax \let\chapterlabel=\relax
\let\sectionlabel=\relax \equanumber=-1 }
\def\titlestyle#1{\par\begingroup \interlinepenalty=9999
     \leftskip=0.02\hsize plus 0.23\hsize minus 0.02\hsize
     \rightskip=\leftskip \parfillskip=0pt
     \hyphenpenalty=9000 \exhyphenpenalty=9000
     \tolerance=9999 \pretolerance=9000
     \spaceskip=0.333em \xspaceskip=0.5em
     \iftwelv@\fourteenpoint\else\twelvepoint\fi
   \noindent #1\par\endgroup }
\def\spacecheck#1{\dimen@=\pagegoal\advance\dimen@ by -\pagetotal
   \ifdim\dimen@<#1 \ifdim\dimen@>0pt \vfil\break \fi\fi}
\def\chapter#1{\par \penalty-300 \vskip\chapterskip
   \spacecheck\chapterminspace
   \chapterreset \titlestyle{\chapterlabel \ #1}
   \nobreak\vskip\headskip \penalty 30000
   \wlog{\string\chapter\ \chapterlabel} }

\def\section#1{\par \ifnum\the\lastpenalty=30000\else
   \penalty-200\vskip\sectionskip \spacecheck\sectionminspace\fi
   \wlog{\string\section\ \chapterlabel \the\sectionnumber}
   \global\advance\sectionnumber by 1  \noindent
   {\caps\enspace\chapterlabel \sectionlabel #1}\par
   \nobreak\vskip\headskip \penalty 30000 }
\def\subsection#1{\par
   \ifnum\the\lastpenalty=30000\else \penalty-100\smallskip \fi
   \noindent\undertext{#1}\enspace \vadjust{\penalty5000}}

\def\undertext#1{\vtop{\hbox{#1}\kern 1pt \hrule}}
\def\ack{\par\penalty-100\medskip \spacecheck\sectionminspace
   \line{\fourteenrm\hfil ACKNOWLEDGEMENTS\hfil}\nobreak\vskip\headskip }
\def\APPENDIX#1#2{\par\penalty-300\vskip\chapterskip
   \spacecheck\chapterminspace \chapterreset \xdef\chapterlabel{#1}
   \titlestyle{APPENDIX #2} \nobreak\vskip\headskip \penalty 30000
   \wlog{\string\Appendix\ \chapterlabel} }
\def\Appendix#1{\APPENDIX{#1}{#1}}
\def\appendix{\APPENDIX{A}{}}
%
%
%
\def\eqname#1{\relax \ifnum\the\equanumber<0
     \xdef#1{{\rm(\number-\equanumber)}}\global\advance\equanumber by -1
    \else \global\advance\equanumber by 1
      \xdef#1{{\rm(\chapterlabel \number\equanumber)}} \fi}
\def\eqinsert#1{\noalign{\dimen@=\prevdepth \nointerlineskip
   \setbox0=\hbox to\displaywidth{\hfil #1}
   \vbox to 0pt{\vss\hbox{$\!\box0\!$}\kern-0.5\baselineskip}
   \prevdepth=\dimen@}}
%

%

%

%
%
\def\GENITEM#1;#2{\par \hangafter=0 \hangindent=#1
    \Textindent{$ #2 $}\ignorespaces}
\outer\def\newitem#1=#2;{\gdef#1{\GENITEM #2;}}
\newdimen\itemsize                \itemsize=30pt
\newitem\item=1\itemsize;
\newitem\sitem=1.75\itemsize;     
\newitem\ssitem=2.5\itemsize;     
\outer\def\newlist#1=#2&#3&#4;{\toks0={#2}\toks1={#3}%
   \count255=\escapechar \escapechar=-1
   \alloc@0\list\countdef\insc@unt\listcount     \listcount=0
   \edef#1{\par
      \countdef\listcount=\the\allocationnumber
      \advance\listcount by 1
      \hangafter=0 \hangindent=#4
      \Textindent{\the\toks0{\listcount}\the\toks1}}
   \expandafter\expandafter\expandafter
    \edef\c@t#1{begin}{\par
      \countdef\listcount=\the\allocationnumber \listcount=1
      \hangafter=0 \hangindent=#4
      \Textindent{\the\toks0{\listcount}\the\toks1}}
   \expandafter\expandafter\expandafter
    \edef\c@t#1{con}{\par \hangafter=0 \hangindent=#4 \noindent}
   \escapechar=\count255}
\def\c@t#1#2{\csname\string#1#2\endcsname}
\newlist\point=\Number&.&1.0\itemsize;
\newlist\subpoint=(\alphabetic&)&1.75\itemsize;
\newlist\subsubpoint=(\roman&)&2.5\itemsize;
\newlist\cpoint=\Roman&.&1.0\itemsize;
%

%
%
%
\newcount\referencecount     \referencecount=0
\newif\ifreferenceopen       \newwrite\referencewrite
\newtoks\rw@toks
\def\NPrefmark#1{\atttach{\rm [ #1 ] }}
\let\PRrefmark=\attach
\def\CErefmark#1{\attach{\scriptstyle  #1 ) }}
\def\refmark#1{\relax\ifPhysRev\PRrefmark{#1}\else\NPrefmark{#1}\fi}
\def\crefmark#1{\relax\CErefmark{#1}}
\def\refend{\refmark{\number\referencecount}}
\newcount\lastrefsbegincount \lastrefsbegincount=0
\def\refsend{\refmark{\count255=\referencecount
   \advance\count255 by-\lastrefsbegincount
   \ifcase\count255 \number\referencecount
   \or \number\lastrefsbegincount,\number\referencecount
   \else \number\lastrefsbegincount-\number\referencecount \fi}}
\def\crefsend{\crefmark{\count255=\referencecount
   \advance\count255 by-\lastrefsbegincount
   \ifcase\count255 \number\referencecount
   \or \number\lastrefsbegincount,\number\referencecount
   \else \number\lastrefsbegincount-\number\referencecount \fi}}
\def\refch@ck{\chardef\rw@write=\referencewrite
   \ifreferenceopen \else \referenceopentrue
   \immediate\openout\referencewrite=referenc.texauxil \fi}
%
{\catcode`\^^M=\active 
  \gdef\obeyendofline{\catcode`\^^M\active \let^^M\ }}%
%
{\catcode`\^^M=\active 
  \gdef\ignoreendofline{\catcode`\^^M=5}}
{\obeyendofline\gdef\rw@start#1{\def\t@st{#1} \ifx\t@st\blankend%
\endgroup \@sf \relax \else \ifx\t@st\bl@nkend \endgroup \@sf \relax%
\else \rw@begin#1
\backtotext
\fi \fi } }
{\obeyendofline\gdef\rw@begin#1
{\def\n@xt{#1}\rw@toks={#1}\relax%
\rw@next}}
\def\blankend{}
{\obeylines\gdef\bl@nkend{
}}
\newif\iffirstrefline  \firstreflinetrue
\def\rwr@teswitch{\ifx\n@xt\blankend \let\n@xt=\rw@begin %
 \else\iffirstrefline \global\firstreflinefalse%
\immediate\write\rw@write{\noexpand\obeyendofline \the\rw@toks}%
\let\n@xt=\rw@begin%
      \else\ifx\n@xt\rw@@d \def\n@xt{\immediate\write\rw@write{%
        \noexpand\ignoreendofline}\endgroup \@sf}%
             \else \immediate\write\rw@write{\the\rw@toks}%
             \let\n@xt=\rw@begin\fi\fi \fi}
\def\rw@next{\rwr@teswitch\n@xt}
\def\rw@@d{\backtotext} \let\rw@end=\relax
\let\backtotext=\relax

\newdimen\refindent     \refindent=30pt
\def\refitem#1{\par \hangafter=0 \hangindent=\refindent \Textindent{#1}}
\def\REFNUM#1{\space@ver{}\refch@ck \firstreflinetrue%
 \global\advance\referencecount by 1 \xdef#1{\the\referencecount}}
\def\refnum#1{\space@ver{}\refch@ck \firstreflinetrue%
 \global\advance\referencecount by 1 \xdef#1{\the\referencecount}\refend}

\def\REF#1{\REFNUM#1%
 \immediate\write\referencewrite{%
 \noexpand\refitem{#1.}}%
\begingroup\obeyendofline\rw@start}
\def\ref{\refnum\?%
 \immediate\write\referencewrite{\noexpand\refitem{\?.}}%
\begingroup\obeyendofline\rw@start}
\def\Ref#1{\refnum#1%
 \immediate\write\referencewrite{\noexpand\refitem{#1.}}%
\begingroup\obeyendofline\rw@start}
\def\REFS#1{\REFNUM#1\global\lastrefsbegincount=\referencecount
\immediate\write\referencewrite{\noexpand\refitem{#1.}}%
\begingroup\obeyendofline\rw@start}
\newcount\figurecount     \figurecount=0
\newif\iffigureopen       \newwrite\figurewrite
\def\figch@ck{\chardef\rw@write=\figurewrite \iffigureopen\else
   \immediate\openout\figurewrite=figures.texauxil
   \figureopentrue\fi}
\def\FIGNUM#1{\space@ver{}\figch@ck \firstreflinetrue%
 \global\advance\figurecount by 1 \xdef#1{\the\figurecount}}
\def\FIG#1{\FIGNUM#1
   \immediate\write\figurewrite{\noexpand\refitem{#1.}}%
   \begingroup\obeyendofline\rw@start}
\def\fig{\FIGNUM\? fig.~\?%
\immediate\write\figurewrite{\noexpand\refitem{\?.}}%
\begingroup\obeyendofline\rw@start}
\def\figure{\FIGNUM\? figure~\?
   \immediate\write\figurewrite{\noexpand\refitem{\?.}}%
   \begingroup\obeyendofline\rw@start}
\def\Fig{\FIGNUM\? Fig.~\?%
\immediate\write\figurewrite{\noexpand\refitem{\?.}}%
\begingroup\obeyendofline\rw@start}
\def\Figure{\FIGNUM\? Figure~\?%
\immediate\write\figurewrite{\noexpand\refitem{\?.}}%
\begingroup\obeyendofline\rw@start}
\newcount\tablecount     \tablecount=0
\newif\iftableopen       \newwrite\tablewrite
\def\tabch@ck{\chardef\rw@write=\tablewrite \iftableopen\else
   \immediate\openout\tablewrite=tables.texauxil
   \tableopentrue\fi}
\def\TABNUM#1{\space@ver{}\tabch@ck \firstreflinetrue%
 \global\advance\tablecount by 1 \xdef#1{\the\tablecount}}
\def\TABLE#1{\TABNUM#1
   \immediate\write\tablewrite{\noexpand\refitem{#1.}}%
   \begingroup\obeyendofline\rw@start}
\def\Table{\TABNUM\? Table~\?%
\immediate\write\tablewrite{\noexpand\refitem{\?.}}%
\begingroup\obeyendofline\rw@start}
%

%
%
%
\def\masterreset{\global\pagenumber=1 \global\chapternumber=0
   \ifnum\the\equanumber<0\else \global\equanumber=0\fi
   \global\sectionnumber=0
   \global\referencecount=0 \global\figurecount=0 \global\tablecount=0 }
\def\FRONTPAGE{\ifvoid255\else\vfill\penalty-2000\fi
      \masterreset\global\frontpagetrue
      \global\lastf@@t=0 \global\footsymbolcount=0}

\def\paperstyle{\letterstylefalse\normalspace\papersize}
\def\letterstyle{\letterstyletrue\singlespace\lettersize}
\def\papersize{\hsize=36pc\vsize=50pc\hoffset=0.2truein
               \voffset=0.2truein\skip\footins=\bigskipamount}
\def\lettersize{\hsize=36pc\vsize=50pc\hoffset=0.2truein
    \voffset=.2truein\skip\footins=\smallskipamount \multiply
    \skip\footins by 3 }
\paperstyle   
%
%
\def\MEMO{\letterstyle\FRONTPAGE \letterfrontheadline={\hfil}
    \line{\quad\fourteenrm CERN MEMORANDUM\hfil\twelverm\the\date\quad}
    \medskip \memod@f}

\def\memit@m#1{\smallskip \hangafter=0 \hangindent=1in
      \Textindent{\caps #1}}
\def\memod@f{\xdef\mto{\memit@m{To:}}\xdef\from{\memit@m{From:}}%
     \xdef\topic{\memit@m{Topic:}}\xdef\subject{\memit@m{Subject:}}%
     \xdef\rule{\bigskip\hrule height 1pt\bigskip}}
\memod@f
\newskip\lettertopfil
\lettertopfil = 0pt plus 1.5in minus 0pt
\newskip\letterbottomfil
\letterbottomfil = 0pt plus 2.3in minus 0pt
\newskip\spskip \setbox0\hbox{\ } \spskip=-1\wd0
\def\addressee#1{\medskip\rightline{\the\date\hskip 30pt} \bigskip
   \vskip\lettertopfil
   \ialign to\hsize{\strut ##\hfil\tabskip 0pt plus \hsize \cr #1\crcr}
   \medskip\noindent\hskip\spskip}
\newskip\signatureskip       \signatureskip=40pt
\def\signed#1{\par \penalty 9000 \bigskip \dt@pfalse
  \everycr={\noalign{\ifdt@p\vskip\signatureskip\global\dt@pfalse\fi}}
  \setbox0=\vbox{\singlespace \halign{\tabskip 0pt \strut ##\hfil\cr
   \noalign{\global\dt@ptrue}#1\crcr}}
  \line{\hskip 0.5\hsize minus 0.5\hsize \box0\hfil} \medskip }

\def\endletter{\ifnum\pagenumber=1 \vskip\letterbottomfil\supereject
\else \vfil\supereject \fi}
\newbox\letterb@x
\def\lettertext{\par\unvcopy\letterb@x\par}
\def\multiletter{\setbox\letterb@x=\vbox\bgroup
      \everypar{\vrule height 1\baselineskip depth 0pt width 0pt }
      \singlespace \topskip=\baselineskip }
\def\letterend{\par\egroup}
%
%
%
\newskip\frontpageskip
\newtoks\pubtype
\newtoks\Pubnum
\newtoks\pubnum
\newtoks\pubnu
\newtoks\pubn
\newif\ifp@bblock  \p@bblocktrue
\def\PH@SR@V{\doubl@true \baselineskip=24.1pt plus 0.2pt minus 0.1pt
             \parskip= 3pt plus 2pt minus 1pt }
\def\PHYSREV{\paperstyle\PhysRevtrue\PH@SR@V}
\def\titlepage{\FRONTPAGE\paperstyle\ifPhysRev\PH@SR@V\fi
   \ifp@bblock\p@bblock\fi}
\def\nopubblock{\p@bblockfalse}
\def\endpage{\vfil\break}
\frontpageskip=1\medskipamount plus .5fil
\pubtype={\tensl Preliminary Version}
\Pubnum={$\rm CERN-TH.\the\pubnum $}
\pubnum={0000}
\def\p@bblock{\begingroup \tabskip=\hsize minus \hsize
   \baselineskip=1.5\ht\strutbox \topspace-2\baselineskip
   \halign to\hsize{\strut ##\hfil\tabskip=0pt\crcr
   \the \pubn\cr
   \the \Pubnum\cr
   \the \pubnu\cr
   \the \date\cr}\endgroup}
\def\title#1{\vskip\frontpageskip \titlestyle{#1} \vskip\headskip }
\def\author#1{\vskip\frontpageskip\titlestyle{\twelvecp #1}\nobreak}

\def\address#1{\par\kern 5pt\titlestyle{\twelvepoint\it #1}}
\def\andaddress{\par\kern 5pt \centerline{\sl and} \address}

\def\abstract{\vskip\frontpageskip\centerline{\fourteenrm ABSTRACT}
              \vskip\headskip }

%
%
%

\def\\{\relax\ifmmode\backslash\else$\backslash$\fi}
\def\globaleqnumbers{\relax\ifnum\the\equanumber<0%
\else\global\equanumber=-1\fi}
\def\nextline{\unskip\nobreak\hskip\parfillskip\break}

\def\journal#1&#2(#3){\unskip, \sl #1~\bf #2 \rm (19#3) }
\def\cropen#1{\crcr\noalign{\vskip #1}}
\def\crr{\cropen{10pt}}
\def\topspace{\hrule height 0pt depth 0pt \vskip}

\let\int=\intop         
\def\prop{\mathrel{{\mathchoice{\pr@p\scriptstyle}{\pr@p\scriptstyle}{
                \pr@p\scriptscriptstyle}{\pr@p\scriptscriptstyle} }}}
\def\pr@p#1{\setbox0=\hbox{$\cal #1 \char'103$}
   \hbox{$\cal #1 \char'117$\kern-.4\wd0\box0}}
\def\lsim{\mathrel{\mathpalette\@versim<}}
\def\gsim{\mathrel{\mathpalette\@versim>}}
\def\@versim#1#2{\lower0.2ex\vbox{\baselineskip\z@skip\lineskip\z@skip
  \lineskiplimit\z@\ialign{$\m@th#1\hfil##\hfil$\crcr#2\crcr\sim\crcr}}}
\def\leftrightarrowfill{$\m@th \mathord- \mkern-6mu
        \cleaders\hbox{$\mkern-2mu \mathord- \mkern-2mu$}\hfil
        \mkern-6mu \mathord\leftrightarrow$}
\def\lrover#1{\vbox{\ialign{##\crcr
        \leftrightarrowfill\crcr\noalign{\kern-1pt\nointerlineskip}
        $\hfil\displaystyle{#1}\hfil$\crcr}}}
%
%
%
\let\sec@nt=\sec
\def\sec{\relax\ifmmode\let\n@xt=\sec@nt\else\let\n@xt\section\fi\n@xt}
\def\obsolete#1{\message{Macro \string #1 is obsolete.}}
\def\firstsec#1{\obsolete\firstsec \section{#1}}
\def\firstsubsec#1{\obsolete\firstsubsec \subsection{#1}}
\def\thispage#1{\obsolete\thispage \global\pagenumber=#1\frontpagefalse}
\def\thischapter#1{\obsolete\thischapter \global\chapternumber=#1}
\def\nextequation#1{\obsolete\nextequation \global\equanumber=#1
   \ifnum\the\equanumber>0 \global\advance\equanumber by 1 \fi}
\def\BOXITEM{\afterassigment\B@XITEM\setbox0=}
\def\B@XITEM{\par\hangindent\wd0 \noindent\box0 }
%

%
%

%
%

%
%

%

%

%
\def\to{\rightarrow}
%

%

%
%
%
\def\boxit#1{\vbox{\hrule\hbox{\vrule\kern3pt\vbox{\kern3pt#1\kern3pt}
\kern3pt\vrule}\hrule}}
%
%
%
\catcode`@=12 
%



\def\bold#1{\setbox0=\hbox{$#1$}%
     \kern-.025em\copy0\kern-\wd0
     \kern.05em\copy0\kern-\wd0
     \kern-.025em\raise.0433em\box0 }

\newdimen\mk \mk=1em
\def\LL{\hbox{\lower 0.1\mk\hbox{%
\hbox{\vrule height  0.8\mk width 0.06\mk depth 0pt}%
\kern0.2\mk\raise0.3\mk
\hbox{\vrule height  0.5\mk width 0.06\mk depth 0pt}%
\kern-0.31\mk\raise0.75\mk
\vbox{\hrule height 0.06\mk width 0.3\mk depth 0pt}%
\kern-0.3\mk
\vbox{\hrule height 0.06\mk width 0.6\mk depth 0pt}%
\kern-0.35\mk\raise 0.25\mk
\vbox{\hrule height 0.06\mk width 0.35\mk depth 0pt}%
\kern-0.05\mk
\hbox{\vrule height  0.30\mk width 0.06\mk depth 0pt}%
}}}


\def\ZZ{\hbox{$
\not
\kern0.15em\not
\kern-0.21em\lower0.2em
\vbox{\hrule width 0.52em height 0.06em depth 0pt}
\kern-0.50em\raise0.7em
\vbox{\hrule width 0.52em height 0.06em depth 0pt}
$}}
%
%
\newbox\hdbox%
\newcount\hdrows%
\newcount\multispancount%
\newcount\ncase%
\newcount\ncols
\newcount\nrows%
\newcount\nspan%
\newcount\ntemp%
\newdimen\hdsize%
\newdimen\newhdsize%
\newdimen\parasize%
\newdimen\spreadwidth%
\newdimen\thicksize%
\newdimen\thinsize%
\newdimen\tablewidth%
\newif\ifcentertables%
\newif\ifendsize%
\newif\iffirstrow%
\newif\iftableinfo%
\newtoks\dbt%
\newtoks\hdtks%
\newtoks\savetks%
\newtoks\tableLETtokens%
\newtoks\tabletokens%
\newtoks\widthspec%
%
%
\immediate\write15{%
}%
%
%
\tableinfotrue%
\catcode`\@=11
%
%
\def\tstrut{\vrule height3.1ex depth1.2ex width0pt}%
\def\and{\char`\&}
\def\tablerule{\noalign{\hrule height\thinsize depth0pt}}%
\thicksize=1.5pt
\thinsize=0.6pt
\def\thickrule{\noalign{\hrule height\thicksize depth0pt}}%
\def\ctr#1{\hfil\ #1\hfil}%
%
%
%
%
\tablewidth=-\maxdimen%
\spreadwidth=-\maxdimen%
\def\tabskipglue{0pt plus 1fil minus 1fil}%
%
%
\centertablestrue%
%
%
%
%
\parasize=4in%
\gdef\ARGS{########}
\gdef\headerARGS{####}
\def\@mpersand{&}
{\catcode`\|=13
\gdef\letbarzero{\let|0}
\gdef\letbartab{\def|{&&}}%
\gdef\letvbbar{\let\vb|}%
}
{\catcode`\&=4
\def\ampskip{&\omit\hfil&}
\catcode`\&=13
\let&0
\xdef\letampskip{\def&{\ampskip}}%
\gdef\letnovbamp{\let\novb&\let\tab&}
}
\def\begintable{
   \begingroup%
   \catcode`\|=13\letbartab\letvbbar%
   \catcode`\&=13\letampskip\letnovbamp%
   \def\multispan##1{
      \omit \mscount##1%
      \multiply\mscount\tw@\advance\mscount\m@ne%
      \loop\ifnum\mscount>\@ne \sp@n\repeat%
   }
   \def\|{%
      &\omit\widevline&%
   }%
   \ruledtable
}
\long\def\ruledtable#1\endtable{%
%
%
%
   \offinterlineskip
   \tabskip 0pt
   \def\widevline{\vrule width\thicksize}
   \def\endrow{\@mpersand\omit\hfil\crnorm\@mpersand}%
   \def\crthick{\@mpersand\crnorm\thickrule\@mpersand}%
   \def\crthickneg##1{\@mpersand\crnorm\thickrule
          \noalign{{\skip0=##1\vskip-\skip0}}\@mpersand}%
   \def\crnorule{\@mpersand\crnorm\@mpersand}%
   \def\crnoruleneg##1{\@mpersand\crnorm
          \noalign{{\skip0=##1\vskip-\skip0}}\@mpersand}%
   \let\nr=\crnorule
   \def\endtable{\@mpersand\crnorm\thickrule}%
   \let\crnorm=\cr
%
%
   \edef\cr{\@mpersand\crnorm\tablerule\@mpersand}%
   \def\crneg##1{\@mpersand\crnorm\tablerule
          \noalign{{\skip0=##1\vskip-\skip0}}\@mpersand}%
   \let\ctneg=\crthickneg
   \let\nrneg=\crnoruleneg
   \the\tableLETtokens
%
%
   \tabletokens={&#1}
%
%
   \countROWS\tabletokens\into\nrows%
   \countCOLS\tabletokens\into\ncols%
%
%
   \advance\ncols by -1%
   \divide\ncols by 2%
   \advance\nrows by 1%
%
%
   \iftableinfo %
      \immediate\write16{[Nrows=\the\nrows, Ncols=\the\ncols]}%
   \fi%
%
%
   \ifcentertables
      \ifhmode \par\fi
      \line{
      \hss
   \else %
      \hbox{%
   \fi
      \vbox{%
         \makePREAMBLE{\the\ncols}
         \edef\next{\preamble}
         \let\preamble=\next
         \makeTABLE{\preamble}{\tabletokens}
      }
      \ifcentertables \hss}\else }\fi
   \endgroup
   \tablewidth=-\maxdimen
   \spreadwidth=-\maxdimen
}
\def\makeTABLE#1#2{
   {
   \let\ifmath0
   \let\header0
   \let\multispan0
%
%
   \ncase=0%
   \ifdim\tablewidth>-\maxdimen \ncase=1\fi%
   \ifdim\spreadwidth>-\maxdimen \ncase=2\fi%
   \relax
%
   \ifcase\ncase %
      \widthspec={}%
   \or %
      \widthspec=\expandafter{\expandafter t\expandafter o%
                 \the\tablewidth}%
   \else %
      \widthspec=\expandafter{\expandafter s\expandafter p\expandafter r%
                 \expandafter e\expandafter a\expandafter d%
                 \the\spreadwidth}%
   \fi %
   \xdef\next{
      \halign\the\widthspec{%
      #1
      \noalign{\hrule height\thicksize depth0pt}
      \the#2\endtable
%
      }
   }
   }
   \next
}
\def\makePREAMBLE#1{
   \ncols=#1
   \begingroup
   \let\ARGS=0
   \edef\xtp{\widevline\ARGS\tabskip\tabskipglue%
   &\ctr{\ARGS}\tstrut}
   \advance\ncols by -1
   \loop
      \ifnum\ncols>0 %
      \advance\ncols by -1%
      \edef\xtp{\xtp&\vrule width\thinsize\ARGS&\ctr{\ARGS}}%
   \repeat
   \xdef\preamble{\xtp&\widevline\ARGS\tabskip0pt%
   \crnorm}
   \endgroup
}
\def\countROWS#1\into#2{
   \let\countREGISTER=#2%
   \countREGISTER=0%
   \expandafter\ROWcount\the#1\endcount%
}%
\def\ROWcount{%
   \afterassignment\subROWcount\let\next= %
}%
\def\subROWcount{%
   \ifx\next\endcount %
      \let\next=\relax%
   \else%
      \ncase=0%
      \ifx\next\cr %
         \global\advance\countREGISTER by 1%
         \ncase=0%
      \fi%
      \ifx\next\endrow %
         \global\advance\countREGISTER by 1%
         \ncase=0%
      \fi%
      \ifx\next\crthick %
         \global\advance\countREGISTER by 1%
         \ncase=0%
      \fi%
      \ifx\next\crnorule %
         \global\advance\countREGISTER by 1%
         \ncase=0%
      \fi%
      \ifx\next\crthickneg %
         \global\advance\countREGISTER by 1%
         \ncase=0%
      \fi%
      \ifx\next\crnoruleneg %
         \global\advance\countREGISTER by 1%
         \ncase=0%
      \fi%
      \ifx\next\crneg %
         \global\advance\countREGISTER by 1%
         \ncase=0%
      \fi%
      \ifx\next\header %
         \ncase=1%
      \fi%
      \relax%
      \ifcase\ncase %
         \let\next\ROWcount%
      \or %
         \let\next\argROWskip%
      \else %
      \fi%
   \fi%
   \next%
}
\def\counthdROWS#1\into#2{%
\dvr{10}%
   \let\countREGISTER=#2%
   \countREGISTER=0%
\dvr{11}%
\dvr{13}%
   \expandafter\hdROWcount\the#1\endcount%
\dvr{12}%
}%
\def\hdROWcount{%
   \afterassignment\subhdROWcount\let\next= %
}%
\def\subhdROWcount{%
   \ifx\next\endcount %
      \let\next=\relax%
   \else%
      \ncase=0%
      \ifx\next\cr %
         \global\advance\countREGISTER by 1%
         \ncase=0%
      \fi%
      \ifx\next\endrow %
         \global\advance\countREGISTER by 1%
         \ncase=0%
      \fi%
      \ifx\next\crthick %
         \global\advance\countREGISTER by 1%
         \ncase=0%
      \fi%
      \ifx\next\crnorule %
         \global\advance\countREGISTER by 1%
         \ncase=0%
      \fi%
      \ifx\next\header %
         \ncase=1%
      \fi%
\relax%
      \ifcase\ncase %
         \let\next\hdROWcount%
      \or%
         \let\next\arghdROWskip%
      \else %
      \fi%
   \fi%
   \next%
}%
{\catcode`\|=13\letbartab
\gdef\countCOLS#1\into#2{%
   \let\countREGISTER=#2%
   \global\countREGISTER=0%
   \global\multispancount=0%
   \global\firstrowtrue
   \expandafter\COLcount\the#1\endcount%
   \global\advance\countREGISTER by 3%
   \global\advance\countREGISTER by -\multispancount
}%
\gdef\COLcount{%
   \afterassignment\subCOLcount\let\next= %
}%
{\catcode`\&=13%
\gdef\subCOLcount{%
   \ifx\next\endcount %
      \let\next=\relax%
   \else%
      \ncase=0%
      \iffirstrow
         \ifx\next& %
            \global\advance\countREGISTER by 2%
            \ncase=0%
         \fi%
         \ifx\next\span %
            \global\advance\countREGISTER by 1%
            \ncase=0%
         \fi%
         \ifx\next| %
            \global\advance\countREGISTER by 2%
            \ncase=0%
         \fi
         \ifx\next\|
            \global\advance\countREGISTER by 2%
            \ncase=0%
         \fi
         \ifx\next\multispan
            \ncase=1%
            \global\advance\multispancount by 1%
         \fi
         \ifx\next\header
            \ncase=2%
         \fi
         \ifx\next\cr       \global\firstrowfalse \fi
         \ifx\next\endrow   \global\firstrowfalse \fi
         \ifx\next\crthick  \global\firstrowfalse \fi
         \ifx\next\crnorule \global\firstrowfalse \fi
         \ifx\next\crnoruleneg \global\firstrowfalse \fi
         \ifx\next\crthickneg  \global\firstrowfalse \fi
         \ifx\next\crneg       \global\firstrowfalse \fi
      \fi
\relax
      \ifcase\ncase %
         \let\next\COLcount%
      \or %
         \let\next\spancount%
      \or %
         \let\next\argCOLskip%
      \else %
      \fi %
   \fi%
   \next%
}%
\gdef\argROWskip#1{%
   \let\next\ROWcount \next%
}
\gdef\arghdROWskip#1{%
   \let\next\ROWcount \next%
}
\gdef\argCOLskip#1{%
   \let\next\COLcount \next%
}
}
}
\def\spancount#1{
   \nspan=#1\multiply\nspan by 2\advance\nspan by -1%
   \global\advance \countREGISTER by \nspan
   \let\next\COLcount \next}%
\def\dvr#1{\relax}%
\def\header#1{%
\dvr{1}{\let\cr=\@mpersand%
\hdtks={#1}%
\counthdROWS\hdtks\into\hdrows%
\advance\hdrows by 1%
\ifnum\hdrows=0 \hdrows=1 \fi%
\dvr{5}\makehdPREAMBLE{\the\hdrows}%
\dvr{6}\getHDdimen{#1}%
{\parindent=0pt\hsize=\hdsize{\let\ifmath0%
\xdef\next{\valign{\headerpreamble #1\crnorm}}}\dvr{7}\next\dvr{8}%
}%
}\dvr{2}}
\def\makehdPREAMBLE#1{
\dvr{3}%
\hdrows=#1
{
\let\headerARGS=0%
\let\cr=\crnorm%
\edef\xtp{\vfil\hfil\hbox{\headerARGS}\hfil\vfil}%
\advance\hdrows by -1
\loop
\ifnum\hdrows>0%
\advance\hdrows by -1%
\edef\xtp{\xtp&\vfil\hfil\hbox{\headerARGS}\hfil\vfil}%
\repeat%
\xdef\headerpreamble{\xtp\crcr}%
}
\dvr{4}}
\def\getHDdimen#1{%
\hdsize=0pt%
\getsize#1\cr\end\cr%
}
\def\getsize#1\cr{%
\endsizefalse\savetks={#1}%
\expandafter\lookend\the\savetks\cr%
\relax \ifendsize \let\next\relax \else%
\setbox\hdbox=\hbox{#1}\newhdsize=1.0\wd\hdbox%
\ifdim\newhdsize>\hdsize \hdsize=\newhdsize \fi%
\let\next\getsize \fi%
\next%
}%
\def\lookend{\afterassignment\sublookend\let\looknext= }%
\def\sublookend{\relax%
\ifx\looknext\cr %
\let\looknext\relax \else %
   \relax
   \ifx\looknext\end \global\endsizetrue \fi%
   \let\looknext=\lookend%
    \fi \looknext%
}%
%
%
\def\tablelet#1{%
   \tableLETtokens=\expandafter{\the\tableLETtokens #1}%
}%
\catcode`\@=12
%
%
\newskip\zatskip \zatskip=0pt plus0pt minus0pt
\def\matth{\mathsurround=0pt}
\def\lsim{\mathrel{\mathpalette\atversim<}}
\def\gsim{\mathrel{\mathpalette\atversim>}}
\def\atversim#1#2{\lower0.7ex\vbox{\baselineskip\zatskip\lineskip\zatskip
  \lineskiplimit 0pt\ialign{$\matth#1\hfil##\hfil$\crcr#2\crcr\sim\crcr}}}
\referenceminspace=10pc
\hyphenation{brems-strahlung}
\def\NPrefmark#1{\attach{\scriptstyle #1 )}}

\def\to{\rightarrow}

\def\rchi{\raise2pt\hbox{$\chi$}}

\def\psl{p\hskip-6pt/}

\def\zero{0\hskip-6pt/}
\def\Re{{\cal R \mskip-4mu \lower.1ex \hbox{\it e}}}
\def\Im{{\cal I \mskip-5mu \lower.1ex \hbox{\it m}}}

\def\etal{{\it et~al}.}

\def\and{\! + \!}
\def\rlh{\scriptstyle{\rightharpoonup\hskip-8pt{\leftharpoondown}}}

\def\Im{{\cal I}\mskip-5mu\lower.1ex\hbox{\it m}}
\def\harpar{\partial\hskip-8pt\raise9pt\hbox{$\rlh$}}
\def\crc{\crcr\noalign{\vskip -6pt}}
%
%
\newtoks\Pubnumtwo
\newtoks\Pubnumthree
\catcode`@=11
\def\p@bblock{\begingroup\tabskip=\hsize minus\hsize
   \baselineskip=1.5\ht\strutbox\topspace-2\baselineskip
   \halign to \hsize{\strut ##\hfil\tabskip=0pt\crcr
   \the\Pubnum\cr  \the\Pubnumtwo\cr \the \Pubnumthree\cr
   \the\date\cr \the\pubtype\cr}\endgroup}
\catcode`@=12
\Pubnum{\bf FSU-HEP-930322}
\Pubnumtwo{\bf UIUC-HEP-93-01}
\Pubnumthree{\bf DTP/93/14}
\date={March 1993}
\pubtype={}
\titlepage
\doublespace
\title{\fourteenrm Ratios of $W^\pm\gamma$ and $Z\gamma$ Cross Sections:
\break
New Tools in Probing the Weak Boson Sector at the Tevatron}
\normalspace
\vskip .2in
\author{\fourteenrm U.~Baur\rlap,$^1$ S.~Errede\rlap,$^2$ and
J.~Ohnemus$^3$}
\vskip 2.mm
\centerline{\it $^1$Physics Department, Florida State University,
Tallahassee, FL 32306, USA}
\centerline{\it $^2$Physics Department, University of Illinois, Urbana,
IL 61801, USA}
\centerline{\it $^3$Physics Department, University of Durham, DH1 3LE,
England}
\normalspace
\vskip .7in
\abstract
The ratios ${\cal R}_{\gamma ,\ell}=B(Z\to\ell^+\ell^-)\cdot\sigma(Z\gamma)
/\allowbreak B(W\to\ell\nu)\cdot\sigma(W^\pm\gamma)$,
${\cal R}_{\gamma ,\nu}=B(Z\to\bar\nu\nu)\cdot\sigma(Z\gamma)/\allowbreak
B(W\to\ell\nu)\cdot\sigma(W^\pm\gamma)$,
${\cal R}_{W\gamma}=\sigma(W^\pm\gamma)/\allowbreak \sigma(W^\pm)$, and
${\cal R}_{Z\gamma}=\sigma(Z\gamma)/\allowbreak\sigma(Z)$
are studied as tools to probe the electroweak boson self-interactions.
As a function of the minimum photon transverse momentum,
${\cal R}_{\gamma ,\ell}$ and ${\cal R}_{\gamma ,\nu}$ are found to
directly reflect the
radiation zero present in $W^\pm\gamma$ production in the Standard
Model. All four ratios are sensitive to anomalous $WW\gamma$ and/or
$ZZ\gamma/Z\gamma\gamma$ couplings. The sensitivity of the cross
section ratios to the cuts imposed on the final state
particles, as well as the systematic uncertainties resulting from
different parametrizations of parton distribution functions, the choice
of the factorization scale $Q^2$, and from higher order QCD corrections are
explored. Taking into account these uncertainties,
sensitivity limits for anomalous three gauge
boson couplings, based on a measurement of the cross section ratios with
an integrated luminosity of 25~pb$^{-1}$ at the Tevatron, are estimated.
\endpage

\def\PL #1 #2 #3 {Phys. Lett.~{\bf#1}, #2 (#3)}
\def\NP #1 #2 #3 {Nucl. Phys.~{\bf#1}, #2 (#3)}
\def\ZP #1 #2 #3 {Z.~Phys.~{\bf#1}, #2 (#3)}
\def\PR #1 #2 #3 {Phys. Rev.~{\bf#1}, #2 (#3)}
\def\PRD #1 #2 #3 {Phys. Rev.~D {\bf#1}, #2 (#3)}
\def\PP #1 #2 #3 {Phys. Rep.~{\bf#1}, #2 (#3)}
\def\PRL #1 #2 #3 {Phys. Rev.~Lett.~{\bf#1}, #2 (#3)}
%
\REF\BB{U.~Baur and E.~L.~Berger, \PR D41 1476 1990 .}
\REF\BaBe{U.~Baur and E.~L.~Berger, FSU-HEP-921030, CERN-TH.6680/92
preprint, October 1992, to appear in Phys.~Rev.~{\bf D}.}
\REF\Rat{F.~Halzen and K.~Mursula, \PRL 51 857 1983 ;\nextline
K.~Hikasa, \PR D29 1939 1984 ;\nextline
N.~G.~Deshpande \etal, \PRL 54 1757 1985 ;\nextline
A.~D.~Martin, R.~G.~Roberts, and W.~J.~Stirling, \PL 189B 220 1987 ;
\nextline
E.~L.~Berger, F.~Halzen, C.~S.~Kim, and S.~Willenbrock, \PR D40 83 1989 .
}
\REF\Rexp{C.~Albajar \etal\ (UA1 Collaboration), \PL 253B 503 1991 ;
\nextline
J.~Alitti \etal\ (UA2 Collaboration), \PL 276B 365 1992 ;\nextline
F.~Abe \etal\ (CDF Collaboration), \PR D44 29 1991 \ and \PRL 69 28 1992 .}
\REF\Barg{V.~Barger, T.~Han, J.~Ohnemus, and D.~Zeppenfeld,
\PRL 62 1971 1989 ; \PR D40 2888 1989 ; \PR D41 1715 1990 (E).}
\REF\BZ{U.~Baur and D.~Zeppenfeld, \NP B308 127 1988 .}
\REF\JO{J.~Ohnemus, \PR D47 940 1993 .}
\REF\RADZ{Zhu Dongpei, \PR D22 2266 1980 ; \nextline
C.~J.~Goebel, F.~Halzen, and J.~P.~Leveille, \PR D23 2682 1981 ;
\nextline S.~J.~Brodsky and R.~W.~Brown, \PRL 49 966 1982 ; \nextline
R.~W.~Brown, K.~L.~Kowalski, and S.~J.~Brodsky, \PR D28 624 1983 ;
\nextline M.~A.~Samuel, \PR D27 2724 1983 .}
\REF\jjg{F.~A.~Berends \etal\ , \PL 103B 124 1981 ;\nextline
P.~Aurenche \etal\ , \PL 140B 87 1984 \ and \NP B286 553 1987 ;\nextline
V.~Barger, T.~Han, J.~Ohnemus, and D.~Zeppenfeld, \PL 232B 371 1989 .}
\REF\CDF{F.~Abe \etal\ (CDF Collaboration), \PR D45 3921 1992 .}
\REF\priv{H.~Wahl, private communication.}
\REF\John{J.~Womersley, private communication;\nextline
R.~J.~Madaras, FERMILAB-Conf-92/365-E, to appear in the Proceedings of
the ``7th Meeting of the American Physical Society Division of Particles
and Fields (DPF92)'', Fermilab, Batavia, IL, November 1992;\nextline
M.~Cobal, FERMILAB-Conf-92/358-E, to appear in the Proceedings of the
``4th Topical Seminar on the Standard Model and Just Beyond'', San
Miniato, Italy, June~1992.}
\REF\BC{S.~Bethke and S.~Catani, Proceedings of the XXVIIth Rencontre
de Moriond, ``QCD and High Energy Hadronic Interactions'', Les Arcs,
France, March~22~--~28, 1992, p.~203.}
\REF\HMRS{P.~N.~Harriman, A.~D.~Martin, R.~G.~Roberts, and
W.~J.~Stirling, \PR D42 798 1990 .}
\REF\Cort{J.~Cortes, K.~Hagiwara, and F.~Herzog, \NP B278 26 1986 ;
\nextline
J.~Stroughair and C.~Bilchak, Z.~Phys.~{\bf C26}, 415 (1984);
\nextline J.~Gunion, Z.~Kunszt, and M.~Soldate, \PL 163B 389 1985 ;
\nextline J.~Gunion and M.~Soldate, \PR D34 826 1986 ; \nextline
W.~J.~Stirling \etal, \PL 163B 261 1985 .}
\REF\Wpt{F.~Abe \etal\ (CDF Collaboration), \PRL 66 2951 1991 .}
\REF\MRS{A.~D.~Martin, W.~J.~Stirling, and R.~G.~Roberts,
\PR D47 867 1993 .}
\REF\GRV{M.~Gl\"uck, E.~Reya, and A.~Vogt, \ZP C53 127 1992 .}
\REF\MT{J.~Morfin and W.~K.~Tung, \ZP C52 13 1991 .}
\REF\Jeff{J.~F.~Owens, \PL 266B 126 1991 .}
\REF\NMC{P.~Amaudruz \etal\ (NMC Collaboration), \PL 295B 159 1992 .}
\REF\CCFR{S.~R.~Mishra \etal\ (CCFR Collaboration), NEVIS-1459 preprint
(June 1992).}
\REF\Willy{W.~L.~van Neerven and E.~B.~Zijlstra, \NP B382 11 1992 .}
\REF\BR{H.~Baer and M.~H.~Reno, \PR D43 2892 1991 \ and \PR D45 1503 1992 .}
\REF\FNR{S.~Frixione, P.~Nason, and G.~Ridolfi, \NP B383 3 1992 .}
\REF\Rl{F.~Abe \etal\ (CDF Collaboration), \PRL 64 152 1990 .}
\REF\Wmass{F.~Abe \etal\ (CDF Collaboration), \PRL 65 224 1990 \ and
\PR D43 2070 1991 .}
\REF\LHC{The LHC Study Group, Design Study of the Large Hadron Collider,
CERN 91-03, 1991.}
\REF\BHO{U.~Baur, T.~Han, and J.~Ohnemus, in preparation.}
\REF\SDC{E.~L.~Berger \etal\ (SDC Collaboration), SDC Technical Design
Report, SDC-92-201, April~1992.}
\REF\BHL{R.~Barbieri, H.~Harari, and M.~Leurer, \PL 141B 455 1985 .}
\REF\unit{U.~Baur and D.~Zeppenfeld, \PL 201B 383 1988 .}
\REF\Hagi{W.~J.~Marciano and A.~Queijeiro, \PR D33 3449 1986 ;\nextline
F.~Boudjema, K.~Hagiwara, C.~Hamzaoui, and K.~Numata,
\PR D43 2223 1991 .}
\REF\Pet{J.~Alitti \etal\ (UA2 Collaboration), \PL 277B 194 1992 .}
\REF\Benj{D.~Benjamin, talk given at the ``XXVIIIth Rencontre de
Moriond: Electroweak Interactions and Unified Field Theories'', Les
Arcs, France, March~13 --~20, 1993.}
\REF\HHPZ{K.~Hagiwara \etal, \NP B282 253 1987 .}
\REF\Yang{C.~N.~Yang, \PR 77 242 1950 .}
\REF\Jog{J.~M.~Cornwall, D.~N.~Levin, and G.~Tiktopoulos,
\PRL 30 1268 1973 ; \PR D10 1145 1974 ;\nextline
C.~H.~Llewellyn Smith, \PL 46B 233 1973 ;\nextline
S. D. Joglekar, Ann. of Phys. {\bf 83}, 427 (1974).}
\REF\CDFm{F.~Abe \etal\ (CDF Collaboration), \PRL 69 28 1992 .}
\REF\JR{F.~James and M.~Roos, \NP B172 475 1980 .}
\REF\BDV{J.~Bagger, S.~Dawson, and G.~Valencia, FERMILAB-PUB-92/75-T
preprint (revised August 1992).}
\REF\De{A.~De Rujula \etal, \NP B384 31 1992 ;\nextline
P.~Hern\'andez and F.~J.~Vegas, CERN-TH.6670/92, preprint.}
\REF\BL{C.~Burgess and D.~London, \PRL 69 3428 1992 , McGill-92/04,
McGill-92/05 preprints (March 1992).}
\REF\HISZ{K.~Hagiwara, S.~Ishihara, R.~Szalapski, and D.~Zeppenfeld,
\PL 283B 353 1992 , and MAD/PH/737 preprint (March 1993).}
\REF\muon{P.~M\'ery, S.~E.~Moubarik, M.~Perrottet, and F.~M.~Renard,
\ZP C46 229 1990 .}
\REF\PT{M.~E.~Peskin and T.~Takeuchi, \PRL 65 964 1990 \ and
\PR D46 381 1992 .}
\REF\Alt{G.~Altarelli and R.~Barbieri, \PL 253B 161 1991 ; \nextline
G.~Altarelli, R.~Barbieri, and S.~Jadach, \NP B369 3 1992 .}
\REF\Foxl{P.~M\'ery, M.~Perrottet and F.~M.~Renard, \ZP C38 579 1988 .}
\REF\Boud{G.~Gounaris \etal\ , Proceedings of ``$e^+e^-$ Collisions
at 500~GeV: The Physics Potential'' edt. P.~Zerwas, Vol.~B, p.~735;
\nextline
F.~Boudjema, Proceedings of ``$e^+e^-$ Collisions at 500~GeV:
The Physics Potential'' edt. P.~Zerwas, Vol.~B, p.~757.}
\REF\GG{E.~Yehudai, \PR D41 33 1990 \ and \PR D44 3434 1991 ;\nextline
S.~Y.~Choi and F.~Schrempp, \PL 272B 149 1991 ;\nextline
O.~Philipsen, \ZP C54 643 1992 ;\nextline
S.~Godfrey and K.~A.~Peterson, OCIP/C 92-7, preprint (November 1992).}
%
%
\FIG\one{a) The ratio
${\cal R}_{\gamma ,\ell} = B(Z\to\ell^+\ell^-)\cdot\sigma(Z\gamma)/\allowbreak
B(W\to\ell\nu)\cdot\sigma(W^\pm\gamma)$
as a function of the minimum transverse momentum of the photon,
$p_T^{\rm min}(\gamma)$, at the Tevatron for the cuts summarized in
Eqs.~(2.4) -- (2.7) (solid line). The dashed line shows the
corresponding ratio of $Zj$ to $W^\pm j$ cross sections,
${\cal R}_{j,\ell}$ [see Eq.~(2.10)], versus $p_T^{\rm min}(j)$.
The dotted line, finally, gives the result of
${\cal R}_{\gamma ,\ell}$ for $p_T(\ell)$, $\psl_T>25$~GeV, instead of
the value listed in Eq.~(2.4). \nextline
b) Sensitivity of ${\cal R}_{\gamma ,\ell}$ at the Tevatron to the
cuts imposed. The variation of the cross section ratio, normalized to
${\cal R}_{\gamma ,\ell}$ obtained for the cuts
of Eq.~(2.4), is shown versus $p_T^{\rm min}(\gamma)$. Only one cut at a
time is varied.}
\FIG\two{a) The ratio ${\cal R}_{\gamma ,\ell}$
as a function of the minimum weak boson -- photon invariant mass,
$m_{\rm min}$, at the Tevatron for the cuts summarized in
Eqs.~(2.4) -- (2.7). The solid line shows the ratio for the true
$W\gamma$ invariant mass, whereas the dashed line gives the result if
both solutions of the reconstructed longitudinal neutrino momentum are
used with equal probabilities. \nextline
b) Sensitivity of ${\cal R}_{\gamma ,\ell}$ at the Tevatron to the
cuts imposed. The variation of the cross section ratio, normalized to
${\cal R}_{\gamma ,\ell}$ obtained for the cuts
of Eq.~(2.4), is shown versus $m_{\rm min}$. Only one cut at a time is varied.}
\FIG\three{The ratios
${\cal R}_{W\gamma} = \sigma(W^\pm\gamma) /\allowbreak \sigma(W^\pm)$
(solid line) and
${\cal R}_{Z\gamma} = \sigma(Z\gamma)/\allowbreak \sigma(Z)$
(dashed line) a) as a function of the minimum
photon transverse momentum, $p_T^{\rm min}(\gamma)$, and b) as a
function of the minimum weak boson -- photon invariant mass,
$m_{\rm min}$, at the Tevatron. The cuts summarized in
Eqs.~(2.4) -- (2.7) are imposed. }
\FIG\four{Dependence of ${\cal R}_{\gamma ,\ell}$ on the parametrization
of the parton structure functions. The variation
$\Delta {\cal R}_{\gamma ,\ell}$, normalized to
${\cal R}_{\gamma ,\ell}$ obtained with the HMRSB parametrization, is
shown a) versus $p_T^{\rm min}(\gamma)$ and b) versus $m_{\rm min}$ for
five representative parametrizations. The cuts
used are summarized in Eqs.~(2.4) -- (2.7).}
\FIG\five{ Dependence of a) ${\cal R}_{W\gamma}$ and b)
${\cal R}_{Z\gamma}$ on the parametrization of the parton structure
functions. The variation of the cross section ratios is shown versus
$p_T^{\rm min}(\gamma)$ for five representative fits, normalized to the
cross section ratio obtained with the HMRSB parametrization. The cuts
imposed are summarized in Eqs.~(2.4) -- (2.7).}
\FIG\six{ Dependence of a) ${\cal R}_{W\gamma}$, b)
${\cal R}_{Z\gamma}$, and c) ${\cal R}_{\gamma ,\ell}$ on the choice of
the factorization scale $Q^2$ in the parton distribution functions versus
$p_T^{\rm min}(\gamma)$. The variation of the cross section ratios with
$Q^2$ is shown for $Q^2=m_W^2$ (solid lines) and
$Q^2=100\cdot m_W^2$ (dashed lines), normalized to the
cross section ratio obtained with $Q^2=\hat s$. The cuts
used are summarized in Eqs.~(2.4) -- (2.7).}
\FIG\seven{Sensitivity of a) ${\cal R}_{\gamma ,\ell}$ versus
$p_T^{\rm min}(\gamma)$ and b) ${\cal R}_{\gamma ,\ell}$ versus
$m_{\rm min}$ to higher order QCD corrections.
The variation of the cross section ratio, normalized to the result
obtained in the Born [leading log (LL)] approximation, is shown for the full
next-to-leading log QCD corrections (solid lines), and for the zero-jet
requirement of Eq.~(2.14) (dashed lines). The cuts imposed are listed
in Eqs.~(2.12) and~(2.13).}
\FIG\eight{Sensitivity of a) ${\cal R}_{V\gamma}$ versus
$p_T^{\rm min}(\gamma)$ and b) ${\cal R}_{V\gamma}$ versus
$m_{\rm min}$ ($V=W,\,Z$) to higher order QCD corrections.
The variation of the cross section ratio, normalized to the result
obtained in the Born [leading log (LL)] approximation, is shown for the full
next-to-leading log QCD corrections and for the zero-jet
requirement of Eq.~(2.14). The cuts
used are summarized in Eqs.~(2.12) and~(2.13).}
\FIG\nine{The ratio ${\cal R}_{\gamma ,\ell}$ as a
function of the minimum transverse momentum of the photon,
$p_T^{\rm min}(\gamma)$, at the LHC (dashed line) and SSC (solid line)
for the cuts summarized in Eq.~(3.1). The dotted and dash-dotted line
show the corresponding ratio of $Zj$ to $W^\pm j$ cross sections,
${\cal R}_{j,\ell}$, versus $p_T^{\rm min}(j)$.}
\FIG\ten{Feynman rule for the general $V_1\gamma V_2$, $V_1=W,\,Z$,
$V_2=W,\, Z,\,\gamma$ vertex. $e$ is the charge of the proton.}
\FIG\eleven{The inverse cross section ratio ${\cal R}^{-1}_{\gamma ,\ell}$
at the Tevatron a) versus $p_T^{\rm min}(\gamma)$ and b) versus
$m_{\rm min}$. The cuts imposed are listed in Eqs.~(2.4) -- (2.7).
The curves are for the SM (solid), $\Delta\kappa_0=2.6$ (dashed),
and $\lambda_0=1.7$ (dotted). A dipole form factor ($n=2$) with
$\Lambda=750$~GeV is used to obtain the curves for non-standard
couplings. The error bars indicate the expected
statistical errors for an integrated luminosity of 25~pb$^{-1}$ for
$W\to e\nu$ and $Z\to e^+e^-$ decays. Only one $WW\gamma$ coupling is
varied at a time. All $ZZ\gamma$ and $Z\gamma\gamma$ couplings are
assumed to vanish identically. }
\FIG\twelve{The cross section ratio ${\cal R}_{\gamma ,\ell}$
at the Tevatron a) versus $p_T^{\rm min}(\gamma)$ and b) versus
$m_{\rm min}$. The cuts used are summarized in Eqs.~(2.4) -- (2.7).
The curves are for the SM (solid), $h^Z_{30}=1$ (dashed),
and $h^Z_{40}=0.075$ (dotted). For the form factor parameters [see Eq.~(3.9)]
we assume $n=3$ ($n=4$) for $h^Z_{30}$ ($h^Z_{40}$) with $\Lambda=750$~GeV.
The error bars indicate the expected
statistical errors for an integrated luminosity of 25~pb$^{-1}$ for
$W\to e\nu$ and $Z\to e^+e^-$ decays. Only one $ZZ\gamma$ coupling is
varied at a time. Anomalous $WW\gamma$ and $Z\gamma\gamma$ couplings are
assumed to vanish identically. }
\FIG\thirteen{a) The inverse cross section ratio
${\cal R}^{-1}_{\gamma ,\nu}$ at the Tevatron versus
$p_T^{\rm min}(\gamma)$. The curves are for the SM (solid),
$\Delta\kappa_0=2.6$ (dashed), and $\lambda_0=1.7$ (dotted). A dipole
form factor ($n=2$) with $\Lambda=750$~GeV is used to obtain the curves
for non-standard couplings. Only one $WW\gamma$ coupling is
varied at a time. All $ZZ\gamma$ and $Z\gamma\gamma$ couplings are
assumed to vanish identically. \nextline
b) The cross section ratio ${\cal R}_{\gamma ,\nu}$ at the Tevatron versus
$p_T^{\rm min}(\gamma)$. The curves are for the SM (solid), $h^Z_{30}=1$
(dashed), and $h^Z_{40}=0.075$ (dotted). For the form factor parameters
[see Eq.~(3.9)] we assume $n=3$
($n=4$) for $h^Z_{30}$ ($h^Z_{40}$) with $\Lambda=750$~GeV.
Only one $ZZ\gamma$ coupling is
varied at a time. Anomalous $WW\gamma$ and $Z\gamma\gamma$ couplings are
assumed to vanish identically. \nextline
The cuts imposed are summarized in Eqs.~(2.4) and~(2.7).
The error bars indicate the expected
statistical errors for an integrated luminosity of 25~pb$^{-1}$ for
$W\to e\nu$ decays. }
\FIG\fourteen{The inverse cross section ratio
${\cal R}^{-1}_{\gamma ,\nu}$ at the Tevatron versus
$p_T^{\rm min}(\gamma)$. The curves are for the SM (solid),
$\Delta\kappa_0=2.6$, $h^Z_{40}=0.075$ (dashed), and $\lambda_0=1.7$,
$h^Z_{30}=1.5$ (dotted). The cuts imposed are summarized in Eqs.~(2.4)
and~(2.7). For anomalous $WW\gamma$ couplings a dipole form factor
($n=2$) is used. For non-standard $ZZ\gamma$ couplings we assume $n=3$
($n=4$) for $h^Z_{30}$ ($h^Z_{40}$). The form factor scale is assumed
to be $\Lambda=750$~GeV. The error bars indicate the expected
statistical errors for an integrated luminosity of 25~pb$^{-1}$ for
$W\to e\nu$ decays. }
\FIG\fifteen{a) The cross section ratio ${\cal R}_{W\gamma}$ at the
Tevatron versus $p_T^{\rm min}(\gamma)$. The curves are for the SM (solid),
$\Delta\kappa_0=2.6$ (dashed), and $\lambda_0=1.7$ (dotted). A dipole
form factor ($n=2$) with $\Lambda=750$~GeV is used to obtain the curves
for non-standard couplings. Only one $WW\gamma$ coupling is
varied at a time. All $ZZ\gamma$ and $Z\gamma\gamma$ couplings are
assumed to vanish identically. \nextline
b) The cross section ratio ${\cal R}_{Z\gamma}$ at the Tevatron versus
$p_T^{\rm min}(\gamma)$. The curves are for the SM (solid), $h^Z_{30}=1$
(dashed), and $h^Z_{40}=0.075$ (dotted). For the form factor parameters
we assume [see Eq.~(3.9)] $n=3$
($n=4$) for $h^Z_{30}$ ($h^Z_{40}$) with $\Lambda=750$~GeV.
Only one $ZZ\gamma$ coupling is
varied at a time. Anomalous $WW\gamma$ and $Z\gamma\gamma$ couplings are
assumed to vanish identically. \nextline
The cuts imposed are summarized in Eqs.~(2.4) -- (2.7).
The error bars indicate the expected
statistical errors for an integrated luminosity of 25~pb$^{-1}$ for
$W\to e\nu$ and $Z\to e^+e^-$ decays. }
%
\pagenumber=1
\chapter{Introduction}

The present run of the Tevatron $p\bar p$ collider is expected to result
in a substantial increase of the integrated luminosity. The increase in
statistics will make
it possible to observe new reactions such as $W^\pm\gamma$ and $Z\gamma$
production, and to probe previously untested sectors of the Standard
Model (SM) of electroweak interactions, in particular, the vector boson
self-interactions. Within the SM, at tree
level, these self-interactions are completely fixed by the $SU(2)\times
U(1)$ gauge theory structure of the model. Their observation is thus
a crucial test of the model. In contrast to low energy and high precision
experiments at the $Z$ peak, collider experiments offer the possibility of a
direct, and essentially model independent, measurement of the three vector
boson vertices. For a detailed investigation at the Tevatron, based on
differential cross section
distributions, an integrated luminosity of at least 100~pb$^{-1}$ is
required\rlap.\refmark{\BB,\BaBe} For smaller data samples
the total cross section is also useful.

In hadron collider experiments, cross section measurements are usually
plagued by large experimental systematic and theoretical errors.
These errors, however, can often be significantly reduced by considering
ratios of cross sections. A well known example is the ratio
$$ {\cal R}_\ell={\sigma(W^\pm\to\ell^\pm\nu)
\over\sigma(Z\to\ell^+\ell^-)}=
{B(W\to\ell\nu)\cdot\sigma(W^\pm)\over B(Z\to\ell^+\ell^-)
\cdot\sigma(Z)} \eqno (1.1) $$
of the observable $W^\pm$ and $Z$ cross sections\rlap.\refmark{\Rat}
Here, $\ell=e,\,\mu$, $B(W\to\ell\nu)$ and $B(Z\to\ell^+\ell^-)$ denote
the leptonic branching ratios of the $W$ and $Z$ boson, respectively,
and $\sigma(W^\pm)$ [$\sigma(Z)$] is the $W^\pm$ [$Z$] production cross
section in $p\bar p$ collisions. The systematic error of ${\cal R}_\ell$ is
less than half that of the individual
cross sections $B(W\to\ell\nu)\cdot\sigma(W^\pm)$ and
$B(Z\to\ell^+\ell^-)\cdot\sigma(Z)$\rlap.\refmark{\Rexp} Using the SM
expectation for the cross section ratio $\sigma(W^\pm)/\sigma(Z)$
together with information on the leptonic branching ratio of the $Z$
boson from LEP, $B(W\to\ell\nu)$ can be
determined from ${\cal R}_\ell$, in turn, this value of $B(W\to\ell\nu)$
can be translated into a model
independent lower limit on the top quark mass of $m_t>55$~GeV (95\%~CL)\rlap.
\refmark{\Rexp}

It is natural to consider cross section ratios similar to that of
Eq.~(1.1) for $W^\pm\gamma$ and $Z\gamma$ production, and to use them to
extract information on $WW\gamma$, $ZZ\gamma$, and $Z\gamma\gamma$
couplings. Four different ratios can be formed:
$$ \eqalignno{
{\cal R}_{\gamma ,\ell} &= {B(Z\to\ell^+\ell^-)\cdot\sigma(Z\gamma)\over
B(W\to\ell\nu)\cdot\sigma(W^\pm\gamma)}\,, & (1.2)\cr
& & \cr
{\cal R}_{\gamma ,\nu} &= {B(Z\to\bar\nu\nu)\cdot\sigma(Z\gamma)\over
B(W\to\ell\nu)\cdot\sigma(W^\pm\gamma)}\,, & (1.3)\cr
& & \cr
{\cal R}_{W\gamma} &= {B(W\to\ell\nu)\cdot\sigma(W^\pm\gamma)\over
B(W\to\ell\nu)\cdot\sigma(W^\pm)} =
{\sigma(W^\pm\gamma)\over\sigma(W^\pm)}\,, & (1.4)\cr
\noalign{\hbox{and}}
{\cal R}_{Z\gamma} &= {B(Z\to\ell^+\ell^-)\cdot\sigma(Z\gamma)\over
B(Z\to\ell^+\ell^-)\cdot\sigma(Z)} = {\sigma(Z\gamma)\over\sigma(Z)}\,.
& (1.5)\cr } $$
Similar ratios have also been proposed for $W^\pm +n$~jet and $Z+n$~jet,
$n=1\dots 3$, production\rlap.\refmark{\Barg} The $W^\pm\gamma$ and
$Z\gamma$ cross section ratios are related to ${\cal R}_\ell$ of
Eq.~(1.1) through the sum rule
$$ {\cal R}_\ell\cdot{\cal R}_{\gamma ,\ell}={{\cal R}_{Z\gamma}\over
{\cal R}_{W\gamma}}~. \eqno (1.6) $$
Experimentally, the ratios of Eqs.~(1.2) -- (1.5) can be
determined from independent data samples. ${\cal R}_{\gamma ,\ell}$
can be measured from an event sample with at least one isolated, high
transverse momentum electron (muon) and one isolated high $p_T$ photon.
${\cal R}_{\gamma ,\nu}$ can be determined from a data sample extracted
with a missing transverse energy trigger and an additional isolated
hard photon. Finally,
${\cal R}_{W\gamma}$ and ${\cal R}_{Z\gamma}$ can be obtained from the
inclusive sample of $W$ and $Z$ boson candidates, respectively.

Many experimental
uncertainties, for example those associated with lepton and photon detection
efficiencies, or the uncertainty in the integrated luminosity,
are expected to cancel, at least partially, in the cross section ratios.
${\cal R}_{W\gamma}$ and ${\cal R}_{Z\gamma}$ are
independent of the vector boson branching ratios, and thus represent
directly the ratio of $W^\pm\gamma$ to $W^\pm$, and $Z\gamma$ to $Z$,
cross sections. Since the cross section for $W/Z$ production is much
larger than the rate for $W\gamma / Z\gamma$ production, the statistical
error of ${\cal R}_{W\gamma}$ and ${\cal R}_{Z\gamma}$ is expected to
be significantly
smaller than that of ${\cal R}_{\gamma ,\ell}$ and ${\cal R}_{\gamma ,\nu}$.

In this paper we study the theoretical aspects of the cross section
ratios shown in Eqs.~(1.2) -- (1.5). Our calculations are based on
results presented in Refs.~\BB,~\BaBe,~\BZ, and~\JO. Cross sections in
the Born approximation are
obtained by calculating helicity amplitudes for the complete processes
$q\bar q'\to W^\pm\gamma\to\ell^\pm\nu\gamma$, $q\bar q\to
Z\gamma\to\ell^+\ell^-\gamma$, and $q\bar q\to
Z\gamma\to\nu\bar\nu\gamma$, including the effects of timelike photon
exchange diagrams and bremsstrahlung from the final state lepton line.
Finite $W/Z$ width effects, and correlations between the final state
leptons originating from $W/Z$ decay, are also fully incorporated in our
calculations. In contrast, next to leading log QCD corrections to $W\gamma$ and
$Z\gamma$ production are at present only known in the limit of stable,
onshell weak bosons\rlap.\refmark{\JO}

In Section~2 we consider the cross section ratios (1.2) --
(1.5) within the framework of the SM at Tevatron energies.
Experimentally, one measures the ratios
$$\eqalignno{
\widetilde{\cal R}_{\gamma ,\ell} &= {\sigma(\ell^+\ell^-\gamma)
\over\sigma(\ell^\pm\nu\gamma)}\, , & (1.7a) \cr
\widetilde{\cal R}_{\gamma ,\nu} &= {\sigma(\bar\nu\nu\gamma)
\over\sigma(\ell^\pm\nu\gamma)}\, , & (1.7b) \cr
\widetilde{\cal R}_{W\gamma} &= {\sigma(\ell^\pm\nu\gamma)
\over\sigma(\ell^\pm\nu)}\, , & (1.7c) \cr
\noalign{\hbox{and}}
\widetilde{\cal R}_{Z\gamma} &= {\sigma(\ell^+\ell^-\gamma)
\over\sigma(\ell^+\ell^-)}\, , & (1.7d)\cr} $$
rather than ${\cal R}_{\gamma ,\ell}$, ${\cal R}_{\gamma ,\nu}$, and ${\cal
R}_{V\gamma}$ ($V=W,\,Z$) directly. In order to
isolate the cross section ratios of Eqs.~(1.2) -- (1.5), appropriate
cuts have to be imposed in order to suppress the contributions of final
state bremsstrahlung (radiative $W/Z$ decays) to the $\ell\nu\gamma$ and
$\ell\ell\gamma$ final states. These cuts are described in
Section~(2.1), together with other details of our calculation.

In Section~(2.2), the ratios are studied as a function of
the minimum photon transverse momentum, $p_T^{\rm min}(\gamma)$, and the
minimum $V\gamma$ ($V=W,\,Z$) invariant mass, $m_{\rm min}$. As a
function of $p_T^{\rm min}(\gamma)$,
${\cal R}_{\gamma ,\ell}$ and ${\cal R}_{\gamma ,\nu}$ are shown to
directly reflect the
radiation zero which is present in $W\gamma$ production in the
SM\rlap.\refmark{\RADZ} In
Section~2.2 we also investigate how the ratios depend on the cuts
imposed on the final state particles. The systematic and theoretical
uncertainties of the cross section ratios originating from the
parametrization of the parton distribution functions, the choice of the
factorization scale $Q^2$, and higher order QCD corrections are studied
in Section~(2.3). The size of the QCD corrections can be
reduced significantly by imposing a central jet veto cut. The theoretical and
systematic uncertainties to the cross section ratios are found to be
well under control. ${\cal R}_{\gamma ,\ell}$ and ${\cal R}_{\gamma ,
\nu}$ are significantly less sensitive to these uncertainties than
${\cal R}_{W\gamma}$ and ${\cal R}_{Z\gamma}$. The $W^\pm\gamma$
and $Z\gamma$ cross section ratios thus possess the same advantages
which make the ratio of $W$ to $Z$ boson cross sections, Eq.~(1.1), a
powerful tool for probing new physics, {\it e.g.}, the extraction
of a model
independent limit on the top quark mass\rlap.\refmark{\Rat,\Rexp}

In Section~3 we study how non-standard three gauge boson couplings
affect the cross section ratios. We also estimate the sensitivity limits
for anomalous three vector boson couplings which one can hope to achieve
from data accumulated in the current Tevatron run, taking into account
the systematic uncertainties to the ratios. Section~4, finally, contains our
conclusions.

\chapter{Standard Model $W^\pm\gamma$ and $Z\gamma$ Cross Section Ratios}

\section{Preliminaries}

The signal in $p\bar p\to W^\pm\gamma /Z\gamma$ consists of an
isolated high transverse momentum ($p_T$) photon and a $W^\pm$ or $Z$
boson which may decay either hadronically or leptonically. The hadronic
$W$ and $Z$ decays will be difficult to observe due to the QCD
2~jet + $\gamma$ background\rlap.\refmark{\jjg} In the following we therefore
focus on the leptonic decay modes of the weak bosons. The signal for
$W^\pm\gamma$ production is then
$$ p\bar p\to\ell^\pm\psl_T\gamma, \eqno (2.1) $$
where $\ell=e,\,\mu$ (we neglect the $\tau$ decay mode of the $W/Z$) and
the missing transverse momentum $\psl_T$ results from the
nonobservation of the neutrino from the $W$ decay. The signal
for $Z\gamma$ production is
$$ p\bar p\to\ell^+\ell^-\gamma \eqno (2.2) $$
if the $Z$ boson decays into a pair of charged leptons, and
$$ p\bar p\to\psl_T\gamma \eqno(2.3) $$
if the $Z$ boson decays into a pair of neutrinos. Besides the standard
Feynman diagrams for $q\bar q'\to W\gamma$ and $q\bar q\to Z\gamma$,
final state bremsstrahlungs diagrams contribute to (2.1) and (2.2). We
incorporate their effects, together with those from timelike photon
exchange diagrams contributing to (2.2), and finite $W/Z$ width effects,
in our numerical simulations of the lowest order cross sections. All
cross sections and dynamical distributions are evaluated using parton
level Monte Carlo programs.

In order to simulate the finite acceptance of detectors we impose,
unless stated otherwise, the following set of transverse momentum,
pseudorapidity ($\eta$), and separation cuts:
$$\matrix{
p_T(\gamma)> 10~{\rm GeV}, & \qquad & |\eta(\gamma)|<3, \cr
p_T(\ell)> 15~{\rm GeV}, & \qquad & |\eta(\ell)|<3.5, \cr
\psl_T> 15~{\rm GeV},  &  \qquad &  \Delta R(\ell\gamma) > 0.7. \crr}
\eqno (2.4) $$
Here,
$$ \Delta R(\ell\gamma)=\left[\left(\Delta\Phi_{\ell\gamma}\right)^2 +
\left(\Delta\eta_{\ell\gamma}\right)^2\right]^{1/2} \eqno (2.5) $$
is the charged lepton photon separation in the pseudorapidity azimuthal
angle plane. The cuts listed in Eq.~(2.4) approximate the phase space
region covered by the CDF and D$\zero$ detectors at the Tevatron\rlap.
\refmark{\CDF,\priv}

Due to the large separation cut, contributions from
the final state bremsstrahlung (radiative $W/Z$ decay) diagrams to
(2.1) and (2.2) are strongly suppressed. They
can be eliminated almost completely by imposing the following
additional cuts on the
invariant mass of the lepton pair and the $\ell\ell\gamma$ system:
$$ m_{\ell\ell} > 50~{\rm GeV,} \hskip 1.cm m_{\ell\ell\gamma}>100~{\rm
GeV} \eqno (2.6) $$
in reaction~(2.2) (Ref.~\BaBe) and
$$ m_T(\ell\gamma;\psl_T)>90~{\rm GeV} \eqno (2.7) $$
in reaction~(2.1) (Ref.~\BB) where
$$ m^2_T(\ell\gamma;\psl_T)=\left [\left (m^2_{\ell\gamma}+|\bold p_T(\gamma)
+\bold p_T(\ell)|^2\right)^{1/2}+\psl_T\right ]^2-\left |
\bold p_T(\gamma)+\bold p_T(\ell) + \bold{\psl}_T\right |^2 \eqno (2.8) $$
is the square of the cluster transverse mass. In Eq.~(2.8), $m_{\ell\gamma}$
denotes the invariant mass of the $\ell\gamma$ pair.
The cuts listed in Eqs.~(2.6) and~(2.7) ensure that the experimentally
measured cross section ratios of Eq.~(1.7) virtually coincide with
the ratios listed in Eqs.~(1.2) -- (1.5). Therefore, we shall
not discriminate between the two sets of ratios subsequently.

Uncertainties in the energy measurements of the charged leptons and the
photon are taken into account in our numerical simulations by Gaussian
smearing of the particle momenta with
$$ {\sigma\over E}=\cases{ 0.135/\sqrt{E_T}~\oplus~0.02 & for $|\eta| <
1.1$ \crr
0.28/\sqrt{E}~\oplus~0.02 & for $ 1.1<|\eta|<2.4 $ \crr
0.25/\sqrt{E}~\oplus~0.02 & for $2.4<|\eta|<4.2$} \eqno (2.9) $$
corresponding to the CDF detector resolution\rlap.\refmark{\CDF}
In Eq.~(2.9),
$E$ ($E_T$) is the energy (transverse energy) of the particle and
the symbol $\oplus$ signifies that the constant term is added in
quadrature in the resolution. The overall resolution of the
electromagnetic calorimeter of the D$\zero$
detector\refmark{\John} ($\approx 0.15/\sqrt{E}$) is better than that
of the CDF detector. Smearing
effects are therefore less pronounced if the D$\zero$ parametrization
for $\sigma/E$ is used.

The SM parameters used in our calculations are $\alpha=\alpha(m_Z^2)
=1/128$, $\alpha_s(m_Z^2)=0.12$ (Ref.~\BC), $m_Z=91.1$~GeV, and
$\sin^2\theta_W=0.23$. For the parton distribution functions we use
the HMRSB set\refmark{\HMRS} with the scale $Q^2$ given by the parton
center of mass energy squared, $\hat s$, unless stated otherwise.

\section{Basic Properties of the Cross Section Ratios}

Using the results obtained in Refs.~\BB\ and~\BaBe\ it is straightforward
to calculate the cross section ratios (1.2) -- (1.5) within the SM. If
the ratios are considered as a function of the minimum transverse
momentum of the photon, $p_T^{\rm min}(\gamma)$, or the
minimum weak boson -- photon invariant mass, $m_{\rm min}$, they
reflect information carried by the $p_T(\gamma)$ and $m_{V\gamma}$
($V=W,\,Z$) distributions. In the following we shall therefore study the
cross section ratios listed in Eqs.~(1.2) -- (1.5) as a function of
these parameters. We shall also investigate in detail how the ratios are
influenced by the cuts imposed on the final state particles.

Figure~1a shows ${\cal R}_{\gamma ,\ell}$ at the Tevatron as a function of
$p_T^{\rm min}(\gamma)$ for the cuts summarized in Eqs.~(2.4) -- (2.7).
The ratio of $Z\gamma$ to $W^\pm\gamma$ cross sections (solid line) is
seen to increase rapidly with the minimum photon transverse momentum from
${\cal R}_{\gamma ,\ell}\approx 0.3$ at $p_T^{\rm min}(\gamma)=10$~GeV to
${\cal R}_{\gamma ,\ell}\approx 1.2$ at $p_T^{\rm min}(\gamma)=200$~GeV.
This is in sharp contrast to the ratio
%
$$ {\cal R}_{j,\ell} = {B(Z\to\ell^+\ell^-)\cdot\sigma(Zj)\over
B(W\to\ell\nu)\cdot\sigma(W^\pm j)}~, \eqno (2.10) $$
which is shown versus the minimum jet transverse momentum, $p_T^{\rm
min}(j)$, for the same cuts [with the photon replaced by the jet in
Eq.~(2.4)] by the dashed line in Fig.~1a. ${\cal R}_{j,\ell}$ remains in the
range from 0.10 to 0.15 over the whole range of $p_T^{\rm min}(j)$
considered. The slight increase with the minimum jet transverse momentum
is due to the different $x$ behavior of the up- and down-type quark
distribution functions. The ratio of $Zj$ to $W^\pm j$ cross sections
is thus very similar to ${\cal R}_\ell$ [see Eq.~(1.1)], with the $Zj$
production rate suppressed by approximately a factor~10 with respect to
the $W^\pm j$ cross section. On the other hand, the $Z\gamma$
production rate is at most a factor~3 smaller than the $W^\pm\gamma$
cross section in the SM. At large photon transverse momenta,
the rates for $W^\pm\gamma$ and $Z\gamma$ production are similar in
magnitude.

The enhancement of the $Z\gamma$ cross section relative to the
$W^\pm\gamma$ production rate can be understood as a consequence of
the radiation zero present in the SM $q\bar q'\to W\gamma$ matrix
elements\rlap,\refmark{\RADZ} which suppresses $W\gamma$ production.
For $u\bar d\to W^+\gamma$ ($d\bar u\to W^-\gamma$) all contributing
helicity amplitudes vanish for $\cos\Theta=-1/3$ (+1/3), where
$\Theta$ is the angle between the quark and the photon in the parton
center of mass frame.
As a result, the photon rapidity
distribution, $d\sigma/dy^*_\gamma$, in the $W\gamma$ rest frame develops a
dip at zero rapidity when one sums over the $W$
charges\rlap,\refmark{\BB,\BZ} thus reducing the cross section in the
central rapidity region. In contrast,
there is no radiation zero present in $Z\gamma$ production, and
the $y^*_\gamma$ distribution peaks at $y^*_\gamma=0$ for $q\bar q\to
Z\gamma$. For increasing photon transverse momenta, events become more
central in rapidity. The reduction of the
$W^\pm\gamma$ cross section for small rapidities originating from the
radiation zero thus becomes more pronounced at high $p_T(\gamma)$.
This causes the photon transverse momentum distribution of $q\bar q'\to
W^\pm\gamma$ to fall significantly faster than the $p_T(\gamma)$
spectrum of $q\bar q\to Z\gamma$,
which immediately translates into a sharp increase of
${\cal R}_{\gamma ,\ell}$ with $p_T^{\rm min}(\gamma)$.

As mentioned before, the cuts of Eqs.~(2.4) -- (2.7) have been used in
order to obtain ${\cal R}_{\gamma ,\ell}$ shown in Fig.~1a. It is
important to know how the slope of ${\cal R}_{\gamma ,\ell}$ versus
$p_T^{\rm min}(\gamma)$ changes if the geometrical acceptances are
varied. In Fig.~1b we
display the variation of the cross section ratio, normalized to the
ratio obtained with the cuts of Eqs.~(2.4) -- (2.7), $\Delta
{\cal R}_{\gamma ,\ell}/{\cal R}_{\gamma ,\ell}$,
if these cuts are changed. Only one parameter is varied at a time. The
sensitivity of ${\cal R}_{\gamma ,\ell}$ to the cuts imposed in general
decreases for increasing values of $p_T^{\rm min}(\gamma)$. Due to the
radiation zero, the $W^\pm\gamma$ cross section is reduced more
significantly than the $Z\gamma$ production rate, and
${\cal R}_{\gamma ,\ell}$ increases, if the photon is required to be more
central. This is illustrated by the solid line in Fig.~1b, which shows the
variation of ${\cal R}_{\gamma ,\ell}$ if the pseudorapidity cut is
changed
from $|\eta(\gamma)|<3$ to $|\eta(\gamma)|<1$. The shoulder in the region
between $p_T^{\rm min}(\gamma)\approx 30$~GeV and $p_T^{\rm min}(\gamma)
\approx 70$~GeV can also be traced back to the radiation zero. For
small values of the photon transverse momentum, the $\eta(\gamma)$
distribution is very flat in the $W^\pm\gamma$ case. At large
$p_T(\gamma)$, the photon rapidity
spectrum develops a slight dip at $\eta(\gamma)=0$ qualitatively similar
to that in $d\sigma/dy^*_\gamma$. This leads to a shoulder in $\Delta
{\cal R}_{\gamma ,\ell}/{\cal R}_{\gamma ,\ell}$ if the photon rapidity cut
is reduced from $|\eta(\gamma)|<3$ to $|\eta(\gamma)|<1$. If the photon
rapidity
range is reduced even further, this shoulder progresses into a local
maximum, located at $p_T^{\rm min}(\gamma)\approx 50$~GeV. On the other
hand, a more stringent rapidity cut on the charged lepton pseudorapidity of
$|\eta(\ell)|<2$ slightly reduces the cross section ratio (dashed line).
Changes in the lepton photon separation affect ${\cal R}_{\gamma ,\ell}$ very
little, as demonstrated by the dot-dashed line in Fig.~1b.

The dotted line in Fig.~1b, finally, shows the effect of increasing the
$p_T(\ell)$ and $\psl_T$ cuts
from 15~GeV to 25~GeV. It exhibits an interesting structure in the
region around $p_T^{\rm min}(\gamma)=m_W/2\approx 40$~GeV, where $m_W$ is
the $W$ boson mass, which originates from the difference in the
coupling of the leptons to $W$ and $Z$ bosons, and the Jacobian peak
in the lepton $p_T$ distribution. Due to the $V-A$ coupling
of the leptons to the $W$ boson, the charged lepton tends to be
emitted in the direction of the parent $W$, thus picking up most of its
momentum. Hence, the $p_T(\ell)$ distribution is significantly harder
than the $\psl_T$ spectrum in $W^\pm\gamma$ production, whereas the
transverse momentum distributions
of the leptons in $Z\gamma$ production, as a result of the almost pure
axial vector coupling of the charged leptons to the $Z$ boson, almost
coincide. Increasing
the $\psl_T$ and $p_T(\ell)$ cut from 15~GeV to 25~GeV therefore reduces
the $W^\pm\gamma$ cross section more than the $Z\gamma$ production rate,
leading to an increase of ${\cal R}_{\gamma ,\ell}$.
In the region $p_T(\gamma)\gsim m_W/2$, the photon tends to
recoil against (one of) the charged lepton(s). Because of the Jacobian
peak in the $p_T(\ell)$ distribution, the sensitivity of ${\cal
R}_{\gamma ,\ell}$ is strongly enhanced around $p_T^{\rm min}(\gamma)
=40$~GeV. In the cross section ratio, the effect described above
leads to a rather well defined kink in ${\cal R}_{\gamma, \ell}$ versus
the minimum photon transverse momentum at $p_T^{\rm min}(\gamma)\approx
m_W/2$, as demonstrated by the dotted line in Fig.~1a.
At large values of $p_T^{\rm min}(\gamma)$, ${\cal R}_{\gamma ,\ell}$ is almost
independent of the cuts imposed on the final state fermions. This
ensures that the steep rise of ${\cal R}_{\gamma ,\ell}$
with $p_T^{\rm min}(\gamma)$ is not an artifact of the specific set of cuts
applied.

Although we have varied only one cut at a time, the curves in Fig.~1b
correctly reflect the global sensitivity of ${\cal R}_{\gamma ,\ell}$ to
the cuts imposed. For example, changing the lepton rapidity cut from
$|\eta(\ell)|<3.5$ to $|\eta(\ell)|<2$ , and the $p_T(\ell)$ and
$\psl_T$ cut from 15~GeV to 25~GeV at the same time, gives a result for
$\Delta{\cal R}_{\gamma ,\ell}/{\cal R}_{\gamma ,\ell}$ which is quite
similar to that represented by the dotted line in Fig.~1b. For increasing
lepton transverse momenta, events are automatically more central in
rapidity. A more stringent rapidity cut in addition to an increased $p_T$
cut therefore changes the result only slightly.

The cross section ratio ${\cal R}_{\gamma ,\ell}$ as a function of the
minimum invariant mass of the weak boson -- photon system, $m_{\rm min}$,
for Tevatron energies and the cuts of Eqs.~(2.4) -- (2.7) (solid line)
is shown in Fig.~2a. Due to threshold effects originating from the $W/Z$ mass
difference, ${\cal R}_{\gamma ,\ell}$ drops first, before it starts to slowly
rise. For most $W^\pm\gamma$ events with large $W\gamma$ invariant mass,
the photon transverse momentum is fairly small, whereas
$|\eta(\gamma)|$ is large. The radiation zero therefore does not manifest
itself in ${\cal R}_{\gamma ,\ell}$ if the cross section ratio is
considered as a function of $m_{\rm min}$.

At hadron colliders the $W\gamma$ invariant mass cannot be determined
unambiguously because the neutrino from the $W$ decay is not observed. If the
transverse momentum of the neutrino is identified with the missing $p_T$
of a given $W\gamma$ event, the unobservable longitudinal neutrino
momentum can be reconstructed, albeit with a twofold ambiguity, by
imposing the constraint that the neutrino and the charged lepton
four-momenta combine to form the $W$ rest mass\rlap.\refmark{\Cort} On
an event by event basis it is impossible to determine which of the two
solutions corresponds to the actual neutrino longitudinal momentum. In
the following we therefore use both solutions with equal probability when
we consider cross section ratios as a function of the $W\gamma$
invariant mass. This is the most conservative approach possible. The
cross section ratio ${\cal R}_{\gamma ,\ell}$ for the reconstructed
$W\gamma$ invariant mass is
shown by the dashed line in Fig.~2a. Only in the threshold region are the
ratios for the true and reconstructed mass similar.

Figure~2b displays the variation of ${\cal R}_{\gamma ,\ell}$ versus $m_{\rm
min}$, using the reconstructed $W\gamma$ invariant mass, if the cuts of
Eqs.~(2.4) -- (2.7) are changed, normalized to the cross section ratio
obtained with these cuts. As demonstrated by the dashed and
dash dotted curves, a more stringent rapidity cut
on the charged leptons and a less severe separation cut have little
influence on the cross section ratio. Changes in the transverse momentum
and photon rapidity cuts, on the other hand, have a larger effect. If
the $p_T(\ell)$ and $\psl_T$ cut of Eq.~(2.4) is increased to 25~GeV,
the relative change in ${\cal R}_{\gamma ,\ell}$ grows very rapidly with
$m_{\rm min}$ (dotted line). Increasing the minimum
lepton $p_T$ selects a phase
space region
where the two solutions of the longitudinal neutrino momentum tend to be
closer together, so that ${\cal R}_{\gamma ,\ell}$ resembles more closely
the cross section ratio obtained for the true $W\gamma$ invariant mass.
Reducing the photon rapidity range covered, increases the cross section
ratio by 50~--~70\% (solid line).

The results presented in Figs.~1b and~2b have been based on the lowest order
matrix elements of the contributing processes. As a result, the
$W\gamma$ and $Z\gamma$ system is produced with zero transverse
momentum. Higher order QCD corrections give the $W\gamma /Z\gamma$
system a finite $p_T$, and thus may change how the cross section ratio
is affected when the $p_T(\ell)$ and $\psl_T$ cuts are varied. In order
to take these effects properly into account, a complete calculation of
the $W\gamma /Z\gamma$ transverse momentum distribution, including soft
gluon resummation effects, is needed. At present, such a calculation is
not available. However, one expects that the shapes of the
$W\gamma$ and $Z\gamma$ transverse momentum distributions are similar
to those of the $W$ and $Z$ boson $p_T$ distributions.
To roughly estimate how our predictions may change if
the finite $p_T$ of the weak boson -- photon system is taken into
account, we have recalculated $\Delta {\cal R}_{\gamma ,\ell}/{\cal
R}_{\gamma ,\ell}$, smearing the transverse momentum components
of the final state particles using the experimental $p_T$ distribution
of the $W$ boson\rlap.\refmark{\Wpt} Possible differences in the shapes of
$d\sigma/dp_T(W\gamma)$ and $d\sigma/dp_T(Z\gamma)$, and the sensitivity
to details of the $p_T$ spectrum, are simulated by
using different fits to the observed $W$ transverse momentum
distribution. Each fit, appropriately normalized, is then
identified with one of the transverse momentum distributions.
The non-zero transverse momentum of the $W\gamma /Z\gamma$ system turns
out to shift the dotted
curves in Figs.~1b and~2b by typically a few percent. The shapes of the
curves, however, remain almost unchanged.

So far, we have only considered the ratio of $Z\gamma$ to $W^\pm\gamma$
cross sections for $Z$ decays into charged leptons, ${\cal R}_{\gamma ,
\ell}$. The cuts of Eqs.~(2.6) and~(2.7) efficiently suppress
photon radiation from final state leptons, and for equal photon $p_T$ and
rapidity cuts
$$ {\cal R}_{\gamma ,\nu}\approx {B(Z\to\bar\nu\nu)\over
B(Z\to\ell^+\ell^-)}\cdot{\cal R}_{\gamma ,\ell}\approx 6\cdot
{\cal R}_{\gamma ,\ell}\,.  \eqno (2.11) $$
The basic properties of ${\cal R}_{\gamma ,\nu}$ and ${\cal
R}_{\gamma ,\ell}$ are thus the same. In particular, ${\cal
R}_{\gamma ,\nu}$ also rises steeply with the minimum photon $p_T$, reflecting
the radiation zero present in $W\gamma$ production in the SM.

The lowest order prediction for ${\cal R}_{V\gamma}$ ($V=W,\, Z$) at the
Tevatron is shown in Fig.~3, using the cuts summarized in Eqs.~(2.4) --
(2.7). The solid lines give ${\cal R}_{W\gamma}$, whereas the dashed
curves display the corresponding ratio for the $Z\gamma$ case. In order
to calculate the $Z$ boson cross section, $\sigma(Z)$, in ${\cal
R}_{Z\gamma}$, we
have assumed the lepton pair invariant mass to be in the range $65~{\rm
GeV}<m_{\ell\ell}<115$~GeV. Photon exchange contributions and finite $Z$
width effects are fully included in our calculation. Figure~3a shows the
two ratios versus $p_T^{\rm min}(\gamma)$. Due to the radiation zero
present in the SM, ${\cal R}_{W\gamma}$ is considerably smaller than
${\cal R}_{Z\gamma}$, and drops faster with increasing values of the
minimum photon $p_T$. In Fig.~3b the cross section ratios are plotted
versus the minimum weak boson -- photon invariant mass $m_{\rm min}$. As
a result of the twofold ambiguity in the reconstruction of the
longitudinal neutrino momentum, ${\cal R}_{W\gamma}$ decreases more
slowly with $m_{\rm min}$ than ${\cal R}_{Z\gamma}$.
The shape of ${\cal R}_{V\gamma}$ versus $p_T^{\rm min}(\gamma)$ and
$m_{\rm min}$ changes only very little if the cuts on the final state
leptons are varied. The cross section ratios typically vary by 10 --
30\%. For small values of $p_T^{\rm min}(\gamma)$ and $m_{\rm min}$ the
changes in the cross sections cancel almost exactly in the ratio.

\section{Theoretical and Systematic Uncertainties}

Higher order QCD corrections, and the choice of the parametrization of
the parton distribution functions and the factorization scale $Q^2$, are
the premier sources of uncertainties in the calculation of cross
sections in hadronic collisions. It is therefore vital to
investigate their impact on the cross section ratios (1.2) -- (1.5).
The sensitivity of the ratios to the parametrization of the parton
distribution functions is illustrated in Figs.~4 and~5 for
five representative sets:
the MRSS0 and MRSD-- distributions of Ref.~\MRS, the GRVLO\rlap,
\refmark{\GRV} the MTLO\rlap,\refmark{\MT} and the DO1.1\refmark{\Jeff}
parametrization. The MRSS0 and MRSD-- sets take into account new
NMC\refmark{\NMC} and CCFR\refmark{\CCFR} data which
suggest valence and sea quark distributions at low $x$ which lead to
considerably larger cross sections than previous fits. Figure~4 shows the
variation of ${\cal R}_{\gamma ,\ell}$ versus $p_T^{\rm min}(\gamma)$
(Fig.~4a) and $m_{\rm min}$ (Fig.~4b), normalized to the cross section
ratio obtained with the HMRSB set of distribution functions.
Figure~5 displays $\Delta {\cal R}_{W\gamma}/{\cal
R}_{W\gamma}$ (Fig.~5a) and $\Delta {\cal R}_{Z\gamma}/{\cal
R}_{Z\gamma}$ versus $p_T^{\rm min}(\gamma)$ (Fig.~5b).

Although the $Z\gamma$ and $W^\pm\gamma$ total cross sections vary
individually by up to 25\% with the parametrization of the parton
distributions, ${\cal R}_{\gamma ,\ell}$ is found to be
very stable. For ${\cal R}_{\gamma ,\ell}$ as a function
of $p_T^{\rm min}(\gamma)$ ($m_{\rm min}$), the changes are at most 8\%
(13\%) in magnitude for the parametrizations used (see Fig.~4). ${\cal
R}_{W\gamma}$ and ${\cal R}_{Z\gamma}$ are somewhat more sensitive. Here
the ratios vary by up to 18\% and 12\%, respectively, if considered as a
function of the minimum photon transverse momentum (see Fig.~5). The
variation of ${\cal R}_{V\gamma}$ ($V=W,\, Z$) versus $m_{\rm min}$ with
the parametrization of the parton distribution function is
qualitatively and quantitatively similar to that of ${\cal
R}_{\gamma ,\ell}$, and is therefore not shown.

The dependence of the cross section ratios on the factorization scale
$Q^2$ in the parton distribution functions is illustrated in Fig.~6.
In this figure we show the variation of the cross section
ratios, normalized to the corresponding ratio obtained with $Q^2=\hat s$, for
$Q^2=m_W^2$ (solid lines) and $Q^2=100\cdot m_W^2$ (dashed lines) versus
$p_T^{\rm min}(\gamma)$. By choosing these two rather extreme values, we
obtain a fairly conservative estimate of how strongly the cross section
ratios depend on the choice of $Q^2$. In tree level calculations, one
usually chooses a typical energy scale of the hard scattering process,
such as the parton center of mass energy squared, $\hat s$, for $Q^2$.
If the cross section ratios are calculated
to all orders in $\alpha_s$, the result is expected to be independent of
$Q^2$.

For small values of the minimum transverse momentum of the photon and the
weak boson photon invariant mass, all cross section ratios are quite
insensitive to variations in $Q^2$. At large $p_T^{\rm min}(\gamma)$,
however, the changes can be quite large for ${\cal R}_{W\gamma}$
and ${\cal R}_{Z\gamma}$, as illustrated by the solid lines in Figs.~6a
and~6b. The variations of the individual cross sections, however,
cancel to a very good approximation in ${\cal R}_{\gamma ,\ell}$
(Fig.~6c). For $Q^2=100\cdot m_W^2$, the changes in the cross section ratios
with respect to $Q^2=\hat s$ are always smaller than 10\%. Results
similar to those shown in Fig.~6 are also obtained for
${\cal R}_{W\gamma}$ and ${\cal R}_{Z\gamma}$ as a function of $m_{\rm
min}$. ${\cal R}_{\gamma ,\ell}$ is somewhat more sensitive to the choice
of $Q^2$ if considered as a function of $m_{\rm min}$ than the $Z\gamma$
to $W^\pm\gamma$ cross section ratio versus $p_T^{\rm min}(\gamma)$ shown
in Fig.~6c.

The sensitivity of ${\cal R}_{W\gamma}$ and ${\cal R}_{Z\gamma}$ to the
choice of $Q^2$ is
expected to be reduced if next-to-leading log (NLL) QCD corrections are
taken into account. NLL QCD corrections to $q\bar q\to Z\gamma$ and
$q\bar q'\to W\gamma$ have been calculated recently in the framework of
the SM in the limit of a stable, onshell $W/Z$ boson\rlap.\refmark{\JO}
Naively one
might expect that the cross section ratios of Eqs.~(1.2) -- (1.5) change
very little if higher order QCD corrections are incorporated, similar to
the ratio of $W^\pm$ and $Z$ cross sections, ${\cal R}_\ell$, of
Eq.~(1.1) (Ref.~\Willy). Using the results of Refs.~\JO\
and~\BR, we have investigated the influence of NLL QCD corrections on
the cross section ratios. Our results are shown in Figs.~7
and~8. In order to perform a meaningful comparison, the
cross section for $q\bar q'\to W^\pm\gamma$ and $q\bar q\to Z\gamma$ in
the Born approximation is
also calculated in the limit of a stable, onshell $W/Z$ boson. To
roughly simulate detector response, the following transverse momentum
and rapidity cuts are imposed:
$$ p_T(\gamma)>10~{\rm GeV},\hskip 3.mm |\eta(\gamma)|<1, \hskip 3.mm
{\rm and} \hskip 3.mm |y(V)|<2.5. \eqno (2.12) $$
Here, $y(V)$ ($V=W,\,Z$) is the $W/Z$ rapidity. We also require the
photon to be isolated by imposing a cut on the total hadronic energy in
a cone of size $\Delta R=0.7$ about the direction of the photon of
$$ \sum_{\Delta R<0.7} E_{\hbox{\tenrm had}}<0.15\, E_\gamma, \eqno (2.13) $$
where $E_\gamma$ is the photon energy.
This requirement strongly reduces photon bremsstrahlung from final state
quarks and gluons.

The results shown in Figs.~7 and~8 demonstrate that, in contrast to
${\cal R}_\ell$, the NLL QCD corrections to the cross sections only
partially cancel in ${\cal R}_{\gamma ,\ell}$, ${\cal R}_{W\gamma}$, and
${\cal R}_{Z\gamma}$, in particular when these cross section ratios are
considered as a function of $p_T^{\rm min}(\gamma)$. Since the QCD
corrections tend to wash out the SM radiation zero in $q\bar q'\to
W^\pm\gamma$, ${\cal R}_{W\gamma}$ is significantly more sensitive to
NLL order effects than ${\cal R}_{Z\gamma}$ (see Fig.~8a). At large values of
$p_T^{\rm min}(\gamma)$, ${\cal R}_{W\gamma}$ increases by as much as
40\% if QCD corrections are included. On the other hand,
${\cal R}_{\gamma ,\ell}$ is reduced by typically 15~--~20\%  by
${\cal O}(\alpha_s)$ corrections (solid line in Fig.~7a).

Higher order QCD effects are known to change the shapes of the
$p_T(\gamma)$ and invariant mass distributions in $W\gamma$ and
$Z\gamma$ production\rlap.\refmark{\JO} This effect can be traced to
the quark gluon fusion process $qg\to W\gamma q'$ and $qg\to
Z\gamma q$, which carries an enhancement factor $\log^2(p_T^2(\gamma)/m_V^2)$
($V=W,\,Z$) at large values of $p_T(\gamma)$. This enhancement factor
arises from the kinematic region where the photon is
produced at large transverse momentum and recoils against the quark, which
radiates a soft $W/Z$ which is almost collinear to the
quark\rlap.\refmark{\FNR} The shape of the photon $p_T$ distribution is
therefore significantly affected by higher order QCD corrections, and the
corrections to the cross section ratios as a function of $p_T^{\rm
min}(\gamma)$ depend strongly on the minimum photon $p_T$. Since ${\cal
O}(\alpha_s)$ corrections result in a harder $p_T(\gamma)$ distribution,
the corrections to the cross section ratios grow with $p_T^{\rm
min}(\gamma)$. The shape of the $Z\gamma$ and the reconstructed $W\gamma$
invariant mass distribution, on the other hand, is only slightly
affected by higher order QCD corrections. Away from the
threshold region, the corrections to the cross section ratios as a
function of $m_{\rm min}$ are approximately constant.

{}From the discussion above it is clear that the size of the ${\cal
O}(\alpha_s)$ QCD corrections to the cross section ratios
of Eqs.~(1.2) -- (1.5) can be significantly reduced by vetoing hard jets
in the central rapidity region. Requiring
$$ no~{\rm jets~with}\hskip 1.cm p_T(j)>10~{\rm GeV}, \hskip 1.cm
|\eta(j)|<2.5 \eqno (2.14) $$
in the event, we obtain the results shown by the dashed line
(dotted and dash-dotted lines) in Fig.~7 (Fig.~8). A ``zero-jet''
cut similar to that in Eq.~(2.14) has been imposed in the CDF
measurement of the ratio of $W$ to $Z$ cross sections, ${\cal
R}_\ell$\rlap,\refmark{\Rl} and the $W$ mass
measurement\rlap.\refmark{\Wmass} Imposing the jet
veto of Eq.~(2.14), reduces the corrections to the cross section ratios
from higher order QCD effects to the few percent level in the $p_T^{\rm
min}(\gamma)$ and $m_{\rm min}$ range studied.

The results shown in Figs.~7 and~8 have been
obtained for on-shell $W/Z$ bosons. No qualitative changes to these
results are expected if decay correlations, finite $W/Z$ width effects, and
photon exchange diagrams are taken into account. At present, a
calculation of NLL QCD corrections to both, $W^\pm\gamma$ and $Z\gamma$
production, which fully takes into account these effects, does not exist.

As mentioned before, ${\cal R}_{\gamma ,\nu}$ is approximately
proportional to ${\cal R}_{\gamma ,\ell}$ for the cuts imposed [see
Eq.~(2.11)]. The results shown in Figs.~4a, 6c, and~7a therefore apply
also to ${\cal R}_{\gamma ,\nu}$.

\chapter{Measuring Three Vector Boson Couplings in Cross Section Ratios}

\section{${\cal R}_{\gamma ,\ell}$, ${\cal R}_{\gamma ,\nu}$, and the
Standard Model Radiation Zero}

In Section~2.2 we have seen that the strong increase of the ratios of
$Z\gamma$ to $W^\pm\gamma$ cross sections, ${\cal R}_{\gamma ,\ell}$ and
${\cal R}_{\gamma ,\nu}$, as a
function of the minimum transverse momentum of the photon can
be traced to the radiation zero which is present in the SM $q\bar q'\to
W\gamma$ differential cross section. In the last Section we have shown
that ${\cal R}_{\gamma ,\ell}$ and ${\cal R}_{\gamma ,\nu}$ are quite
insensitive to changes in the parametrization of the parton structure
functions. Furthermore, at tree level the two ratios vary
little with a change of the factorization scale $Q^2$. Finally, when a
central jet veto is imposed, the ${\cal O}(\alpha_s)$ QCD corrections
change ${\cal R}_{\gamma ,\ell}$ and ${\cal R}_{\gamma ,\nu}$ by only
a few percent.

The steep rise of ${\cal R}_{\gamma ,\ell}$ and ${\cal R}_{\gamma ,\nu}$
with $p_T^{\rm min}(\gamma)$ in the framework of the SM as a signal of
the radiation zero,
combined with small systematic and theoretical uncertainties, make these
quantities excellent tools to probe the three vector boson vertices.
As we shall see below, anomalous $WW\gamma$
couplings tend to decrease the two ratios, in particular at
large $p_T^{\rm min}(\gamma)$. Non-standard $ZZ\gamma$ and
$Z\gamma\gamma$ couplings, on the other hand, lead to an increase of
${\cal R}_{\gamma ,\ell}$ and ${\cal R}_{\gamma ,\nu}$ to values much
larger than predicted by
the~SM. In contrast to other quantities which are sensitive to the
radiation zero, ${\cal R}_{\gamma ,\ell}$ is fairly simple to measure
experimentally. The prospects for ${\cal R}_{\gamma ,\nu}$ depend on how
well the $p\bar p\to\gamma\psl_T$ signal can be isolated\rlap.
\refmark{\BaBe} As we have mentioned before, the photon rapidity
distribution, $d\sigma/dy^*_\gamma$, in the $W\gamma$ center of mass system
is a quantity which is sensitive to the radiation zero. The measurement
of $d\sigma/dy^*_\gamma$ is complicated by the fact that the neutrino is not
observed, which leads to a twofold ambiguity in the reconstruction of the
$W\gamma$ center of mass system\rlap.\refmark{\Cort} On an event by
event basis it is impossible to decide which of the two solutions is the
correct one. As a result, the radiation zero is partially washed out. On
the other hand, the measurement of ${\cal R}_{\gamma ,\ell}$ and ${\cal
R}_{\gamma ,\nu}$ versus
$p_T^{\rm min}(\gamma)$ is relatively easy, and essentially involves only
counting the number of $W^\pm\gamma$ and $Z\gamma$ events as a function of the
minimum photon transverse momentum.

The ratio ${\cal R}_{\gamma ,\ell}$ may also be very useful in observing the
radiation zero at the LHC [$pp$ collisions at
$\sqrt{s}=15.4$~TeV (Ref.~\LHC)] and
SSC ($pp$ collisions at $\sqrt{s}=40$~TeV).
At these center of mass energies the
higher order QCD corrections to $W^\pm\gamma$ production completely
obscure the radiation zero in $d\sigma/dy^*_\gamma$ (Ref.~\JO), even
when a rather tight central jet veto is imposed\rlap.\refmark{\BHO} The ${\cal
O}(\alpha_s)$ QCD corrections to ${\cal R}_{\gamma ,\ell}$, on the other
hand, are found to be well under control if one requires that no jets with
$p_T(j)>50$~GeV and $|\eta(j)|<3$ are present in the event.

The tree level prediction of ${\cal R}_{\gamma ,\ell}$ at hadron
supercolliders as a function of the minimum photon transverse momentum
is shown in Fig.~9. To simulate detector response,
we have imposed the following set of cuts:
$$\matrix{ p_T(\gamma)>100~{\rm GeV}, & \qquad & |\eta(\gamma)|<3, \cr
p_T(\ell),~\psl_T>20~{\rm GeV}, & \qquad & |\eta(\ell)|<3, \cr
m_{\ell\ell}>50~{\rm GeV}, & \qquad & \Delta R(\ell\gamma)>0.7. }
\eqno (3.1) $$
Energy mismeasurements in the detector were simulated by Gaussian
smearing of the charged lepton and photon momenta using the
expected resolution of the SDC detector\rlap.
\refmark{\SDC}

At the LHC and SSC, ${\cal R}_{\gamma ,\ell}$ grows with
increasing values of $p_T^{\rm min}(\gamma)$, similar to the situation
encountered for Tevatron energies. Due to the smaller center of mass
energy available at the LHC, ${\cal R}_{\gamma ,\ell}$ rises faster than
at SSC energies (solid line). For example, a minimum photon $p_T$ of
1~TeV at LHC energies corresponds to $p_T^{\rm min}(\gamma)\approx
2.6$~TeV at $\sqrt{s}=40$~TeV. For these values of
$p_T^{\rm min}(\gamma)$, ${\cal R}_{\gamma ,\ell}$ is approximately equal
for both energies. The ratio of $Zj$ to $W^\pm j$ cross
sections, ${\cal R}_{j,\ell}$, on the other hand, stays approximately
constant (${\cal R}_{j,\ell}\approx 0.12$) over the entire range of
$p_T^{\rm min}(j)$ values considered (dotted and dash-dotted lines).
Therefore, ${\cal R}_{\gamma ,\ell}$ reflects the radiation zero also at
supercollider energies.

Compared to ${\cal R}_{V\gamma}$ ($V=W,\, Z$), ${\cal R}_{\gamma ,\ell}$
and ${\cal R}_{\gamma ,\nu}$ have the advantage of reflecting the SM
radiation zero. Moreover, at tree level,
systematic and theoretical errors are significantly smaller for these
cross section ratios than for ${\cal R}_{V\gamma}$. On the other hand,
due to the large total $W$ and $Z$ cross section, statistical errors
are expected to be considerably smaller in ${\cal R}_{V\gamma}$.
Furthermore, cancelations between anomalous $WW\gamma$ and
$ZZ\gamma/Z\gamma\gamma$ couplings may occur in ${\cal R}_{\gamma ,\ell}$
and ${\cal R}_{\gamma ,\nu}$ (see below). This is not possible in ${\cal
R}_{V\gamma}=\sigma(V\gamma)/\sigma(V)$. The various $W^\pm\gamma$ and
$Z\gamma$ cross section ratios listed in Eqs.~(1.2) -- (1.5) therefore
yield complementary information on the structure of three vector boson
vertices.

\section{Probing Three Vector Boson Vertices via Cross Section Ratios}

We shall now discuss the impact of non-standard three vector boson
couplings on the $W^\pm\gamma$ and $Z\gamma$ cross section ratios in
more detail. The couplings of $W$ and $Z$ bosons to quarks and leptons
are assumed to be given by the SM. We shall also assume that there are no
non-standard couplings of the $Z\gamma$ pair to two gluons\rlap.
\refmark{\BHL}
The $W$ and $Z$ bosons entering the Feynman diagrams for $q\bar q'\to
W\gamma$ and $q\bar q\to Z\gamma$ couple to essentially massless
fermions, which ensures that effectively $\partial_\mu V^\mu=0$ ($V=W,\,
Z$). This together with gauge invariance of the on-shell photon restricts
the tensor structure of the $WW\gamma$, $ZZ\gamma$, and $Z\gamma\gamma$
vertex to allow just four free parameters. The $WW\gamma$ vertex
function for the process $q\bar q'\to W^\pm\gamma$ is then given
by\refmark{\unit} (see Fig.~10 for notation)
$$ \eqalign{
\Gamma^{\alpha\beta\mu}_{W\gamma W}(q_1,q_2,P) = \mp{1\over 2} \biggl\{
 & (1+\kappa)(q_1-q_2)^\mu g^{\alpha\beta} + {\lambda\over m_W^2}\,
(q_1-q_2)^\mu\,(P^2g^{\alpha\beta}-2P^\alpha P^\beta) \crr
 & - 4\,P^\beta g^{\mu\alpha} +2\, (1+\kappa+\lambda)\,P^\alpha
 g^{\mu\beta} + 2\,
(\tilde\kappa+\tilde\lambda)\,\epsilon^{\mu\alpha\beta\rho}q_{2\rho} \crr
 & +{\tilde\lambda\over m_W^2}\, (q_1-q_2)
 ^\mu\epsilon^{\alpha\beta\rho\sigma}P_\rho\,(q_1-q_2)_\sigma\biggr\}~.
 \crr} \eqno (3.2) $$
The parameters $\kappa$ ($\tilde\kappa$) and $\lambda$ ($\tilde\lambda$)
are related to the magnetic (electric) dipole moment $\mu_W$ ($d_W$) and
the electric (magnetic) quadrupole moment $Q_W$ ($\widetilde Q_W$) of
the $W$ boson by:
$$\eqalignno{\mu_W &= {e\over 2M_W} \left( 1 + \kappa + \lambda
\right)\>, &(3.3a) \cr
Q_W &= -{e\over m^2_W} (\kappa -\lambda) \>, &(3.3b) \cr
d_W &= {e\over 2m_W} (\tilde\kappa + \tilde \lambda)
\>, &(3.3c)\cr
\widetilde Q_W &= -{e\over m^2_W} (\tilde\kappa - \tilde\lambda)
\>.&(3.3d) \cr}$$
While the
$\kappa$ and $\lambda$ terms do not violate any discrete symmetries, the
$\tilde\kappa$ and $\tilde\lambda$ terms are $P$ odd and $CP$ violating.
Within the SM, at tree level,
$$\eqalign{\kappa =1 \>,~~~~& \lambda =0 \>,  \crr
\tilde\kappa =0 \>,~~~~& \tilde\lambda =0 \>.  \cr} \eqno(3.4) $$
The $CP$ violating couplings $\tilde\kappa$ and $\tilde\lambda$ are
constrained by measurements of
the electric dipole moment of the neutron to be smaller
than $\sim 10^{-3}$ in magnitude\rlap.\refmark{\Hagi} Therefore, they will
not be discussed subsequently. The $CP$ conserving couplings
$\kappa$ and $\lambda$ have
been measured recently by the UA2 Collaboration in the process $p\bar
p\to e^\pm\nu\gamma X$ at the CERN $p\bar p$ collider\rlap:
\refmark{\Pet}
$$ \kappa=1\matrix{+2.6\crc -2.2}~~({\rm for}~\lambda=0)\>,\hskip 1.
cm\lambda=0\matrix{+1.7\crc -1.8}~~({\rm for}~\kappa=1)\>, \eqno (3.5) $$
at the 68\% CL. The analysis of the 1988-89 CDF $W\gamma$ (and
$Z\gamma$) data is still in progress\rlap.\refmark{\Benj}

The most general anomalous $Z\gamma Z$ vertex function (see Fig.~10 for
notation) is given by\refmark{\HHPZ}
$$ \eqalign{
\Gamma^{\alpha\beta\mu}_{Z\gamma Z}(q_1,q_2,P) = {P^2-q_1^2\over m_Z^2}~
& \Biggl\{h_1^Z\left(q_2^\mu g^{\alpha\beta}-q_2^\alpha
g^{\mu\beta}\right) \crr
& + {h_2^Z\over m_Z^2}\,P^\alpha\left(\left(P\cdot q_2\right)
g^{\mu\beta}-q_2^\mu P^\beta\right) \crr
& + h_3^Z\epsilon^{\mu\alpha\beta\rho}q_{2\rho} \crr
& + {h_4^Z\over m_Z^2}\,P^\alpha\epsilon^{\mu\beta\rho\sigma}P_\rho
q_{2\sigma}\Biggr\}~,\crr } \eqno (3.6) $$
where $m_Z$ is the $Z$ boson mass. The most general $Z\gamma\gamma$
vertex function can be obtained from Eq.~(3.6) by the following
replacements:
$$ {P^2-q_1^2\over m_Z^2}~\to ~{P^2\over m_Z^2}\hskip 1.cm {\rm
and}\hskip 1.cm h_i^Z\to h_i^\gamma,~~i=1\dots 4. \eqno (3.7) $$
Terms proportional to $P^\mu$ and $q_1^\alpha$ have been omitted in
Eq.~(3.6) since they do not contribute to the cross section.
The overall factor $(P^2-q_1^2)$ in Eq.~(3.6)
is a result of Bose symmetry, whereas the factor $P^2$
in the $Z\gamma\gamma$ vertex function originates from electromagnetic
gauge invariance. As a result the $Z\gamma\gamma$ vertex function vanishes
identically if both photons are onshell\rlap.\refmark{\Yang}
All $ZZ\gamma$ and $Z\gamma\gamma$ couplings are $C$ odd; $h_1^V$
and $h_2^V$ ($V=Z,\gamma$) violate $CP$.
Combinations of $h_3^V$ ($h_1^V$) and $h_4^V$ ($h_2^V$) correspond to
the electric (magnetic) dipole and magnetic (electric) quadrupole
transition moment of the $Z$ boson.  At tree level in the SM, all
couplings $h_i^V$ vanish. Presently, there are no limits on $h^V_i$ from
hadron collider experiments. LEP~I data give only very loose constraints
of $h^V_i\sim{\cal O}(10-100)$ (Ref.~\BaBe).
Without loss of generality we have chosen the overall
$WW\gamma$, $ZZ\gamma$, and $Z\gamma\gamma$ coupling constant to be,
$$ g_{WW\gamma}=g_{ZZ\gamma}=g_{Z\gamma\gamma}=e\, , \eqno (3.8) $$
where $e$ is the charge of the proton.

Tree level unitarity restricts the $WW\gamma$, $ZZ\gamma$, and
$Z\gamma\gamma$ couplings
uniquely to their SM values at asymptotically high energies\rlap.
\refmark{\Jog} This implies that the $WW\gamma$ and $Z\gamma V$
couplings $a=\kappa-1,\dots,\tilde\lambda$ and $h_i^V$ have
to be described by form factors $a(q_1^2,q_2^2,P^2)$ and
$h_i^V(q_1^2,q_2^2,P^2)$ which vanish when $q_1^2$, $q_2^2$, or $P^2$
becomes large. Following Refs.~\BaBe\ and~\BZ, we shall use generalized
dipole form factors of the form
$$ \eqalignno{
a(m_W^2,0,\hat s) = & {a_0\over \left(1+\hat s/\Lambda^2\right)^n}~,
 & (3.9a)\cr
\noalign{\hbox{and}}
h_i^V(m_Z^2,0,\hat s) = & {h_{i0}^V\over \left(1+\hat s/\Lambda^2\right)
^n}~. & (3.9b)} $$
In order to guarantee unitarity, $n$ must satisfy
$n>1/2$ for $a=\Delta\kappa=\kappa-1,\,
\tilde\kappa$, $n>1$ for $a=\lambda,\,\tilde\lambda$ (Ref.~\BZ),
and $n>3/2$ ($n>5/2$) for $h^V_{1,3}$ ($h^V_{2,4}$) (Ref.~\BaBe).
In Eq.~(3.9) $\Lambda$ represents the scale at which new physics becomes
important in the weak boson sector. In the following, we chose
$\Lambda=750$~GeV, $n=2$ for $WW\gamma$ couplings, and $n=3$ ($n=4$) for
$h^V_{1,3}$ ($h^V_{2,4}$).

The influence of anomalous $WW\gamma$ couplings on the ratio of
$Z\gamma$ to $W^\pm\gamma$ cross sections is shown in Fig.~11 for the
cuts summarized in Eqs.~(2.4) -- (2.7). For
presentational reasons we display the inverse cross section ratio
$${\cal R}^{-1}_{\gamma ,\ell}={B(W\to\ell\nu)\cdot\sigma(W^\pm\gamma)
\over B(Z\to\ell^+\ell^-)\cdot\sigma(Z\gamma)}\,. \eqno(3.10) $$
The solid curves show the SM result. The error
bars indicate the statistical errors, corresponding to the 68.3\%
confidence level (CL) interval, expected for an integrated
luminosity of $\int\!{\cal L}dt=25$~pb$^{-1}$ and considering only
$W\to e\nu$ and $Z\to e^+e^-$ decays. If the muon final states of
the weak boson decays are taken into account as well, the statistical
errors may be significantly reduced. Details depend strongly on the
rapidity coverage for muons, which is quite different for
CDF\refmark{\CDFm} and D$\zero$\rlap.\refmark{\John} When estimating the
errors of the cross section ratios, care must be taken for large values
of $p_T^{\rm min}(\gamma)$ and $m_{\rm min}$ where the number of events
in both the numerator and denominator can be very small. To estimate the
statistical errors in these regions, we have used the method described in
Ref.~\JR. For an integrated luminosity of $\int\!{\cal
L}dt=100$~pb$^{-1}$, as foreseen by the end of 1994, the error bars in
Fig.~11, and all subsequent figures are reduced by a factor~1.5 --~2.
The dashed and dotted curves
show ${\cal R}_{\gamma ,\ell}$ for $\Delta\kappa_0=2.6$ and
$\lambda_0=1.7$, the present UA2 68\%~CL limits on the $CP$ conserving
$WW\gamma$ couplings\rlap.\refmark{\Pet} Only one coupling is varied at a
time. For the form factor parameters used ($n=2$ and $\Lambda=750$~GeV),
the values of the two couplings are about a factor~5 and~4 below
the unitarity bound, respectively\rlap.\refmark{\BZ} The anomalous
$ZZ\gamma$ and $Z\gamma\gamma$ couplings, $h_{i0}^V$, are assumed to be
zero in Fig.~11. All numerical results shown in this Section are
obtained using the tree level calculations of Refs.~\BB\ and~\BaBe.

Since the anomalous terms in the helicity amplitudes grow like
$\sqrt{\hat s}/m_W$ for $\Delta\kappa$ and $\hat s/m_W^2$ for $\lambda$,
non-standard $WW\gamma$ couplings lead to an excess of events at large
values of the photon transverse momentum and the $W\gamma$ invariant mass.
As a result, ${\cal R}^{-1}_{\gamma ,\ell}$ is
larger than in the SM if anomalous $WW\gamma$ couplings are present. Due
to the radiation zero one expects ${\cal R}^{-1}_{\gamma ,\ell}$ to
fall with increasing $p_T^{\rm min}(\gamma)$ in the SM (see Fig.~11a).
For anomalous couplings, on the other hand, the inverse cross section
ratio rises very rapidly with the minimum photon transverse momentum.

Fig.~11 shows that it should be possible to measure ${\cal
R}^{-1}_{\gamma ,\ell}$ for
minimum photon transverse momenta up to 40~GeV, and
values of $m_{\rm min}$ up to 200~GeV, with 25~pb$^{-1}$. Comparing
Fig.~11a and~11b it is obvious that ${\cal R}^{-1}_{\gamma ,\ell}$ as a
function of
$p_T^{\rm min}(\gamma)$ is more sensitive to anomalous couplings than the
inverse cross section ratio versus $m_{\rm min}$. The reduced
sensitivity in ${\cal R}^{-1}_{\gamma ,\ell}$ as a function of the minimum
weak boson photon invariant mass is mostly due to the ambiguity in the
reconstructed longitudinal neutrino momentum, $p_{\nu L}$. As before,
we have used both solutions for $p_{\nu L}$ with equal weight in
Fig.~11b. The sensitivity of ${\cal R}^{-1}_{\gamma ,\ell}$ versus $m_{\rm
min}$ would clearly improve if one could discriminate between the two
solutions on a statistical basis. Finally, Fig.~11 demonstrates that, the
UA2 limits on $\Delta\kappa$ and $\lambda$ can be considerably improved
at the Tevatron with an integrated luminosity of 25~pb$^{-1}$.

The impact of anomalous $ZZ\gamma$ couplings on ${\cal R}_{\gamma ,\ell}$
is shown in Fig.~12 for $h^Z_{30}=1$ and $h^Z_{40}=0.075$.
For the form factor parameters used ($n=3$ [$n=4$] for $h^Z_3$
[$h^Z_4$], and $\Lambda=750$~GeV), these values
are approximately a factor~2 below the limit allowed by
unitarity\rlap.\refmark{\BaBe} The $WW\gamma$ and $Z\gamma\gamma$
vertex function are assumed to have SM form in Fig.~12.
For equal coupling strengths, the numerical results obtained for the
$Z\gamma\gamma$ couplings $h_3^\gamma$ and $h_4^\gamma$ are about 20\%
below those obtained for $h_3^Z$ and $h_4^Z$, in the region where
anomalous coupling effects dominate over the SM cross section. Results
for the $CP$ violating couplings $h_{1,2}^V$ ($V=Z,\,\gamma$) are
virtually identical to those obtained for the same values of $h_{3,
4}^V$. Anomalous $ZZ\gamma$ and $Z\gamma\gamma$ couplings are seen to
increase ${\cal R}_{\gamma ,\ell}$ dramatically, especially at large
values of the minimum photon~$p_T$. In contrast to the situation for
anomalous $WW\gamma$
couplings, the sensitivity of ${\cal R}_{\gamma ,\ell}$ to
$ZZ\gamma/Z\gamma\gamma$ couplings is not degraded substantially if
the ratio is considered as a function of $m_{\rm min}$ (see Fig.~12b).

For an integrated luminosity of 25~pb$^{-1}$, the sensitivity of
${\cal R}_{\gamma ,\ell}$ to non-standard three vector boson vertices is
limited mostly by statistical errors. From the results of
Section~2.3 we estimate the systematic errors for ${\cal R}_{\gamma ,\ell}$
to be approximately 10\%. Due to the larger branching ratio of the decay
$Z\to\bar\nu\nu$, the statistical error in the cross section ratio of
Eq.~(1.3), ${\cal R}_{\gamma ,\nu}$, is reduced by a factor~1.4~--~1.7. The
cross section ratio ${\cal R}_{\gamma ,\nu}$ and its inverse are shown in
Fig.~13 as a function of the minimum photon transverse momentum for the
cuts summarized in Eq.~(2.4) and~(2.7). The photon transverse momentum
cut in Fig.~13 has been increased to $p_T(\gamma)>30$~GeV, in order to
suppress backgrounds from $p\bar p\to\gamma j$, with the jet rapidity
outside the range covered by the detector and thus ``faking'' missing
transverse momentum, and two jet production where one of the jets is
misidentified as a photon while the other disappears through the beam
hole\rlap.\refmark{\BaBe} Comparing Fig.~13 with Figs.~11a and~12a, the
increased sensitivity of ${\cal R}_{\gamma ,\nu}$ to anomalous three
vector boson couplings is evident.

So far, we have only varied either $WW\gamma$ or
$ZZ\gamma/Z\gamma\gamma$ couplings. If the three boson vertices contributing to
$W\gamma$ and $Z\gamma$ production simultaneously deviate from the SM,
cancelations may occur between the contributions to
$\sigma(W^\pm\gamma)$ and $\sigma(Z\gamma)$. Couplings
corresponding to operators of different dimension in the effective
Lagrangian have a different high energy behavior, and thus do not cancel
at a substantial level in the cross section ratios. On the other hand,
the effects of $WW\gamma$ and $ZZ\gamma/Z\gamma\gamma$ couplings of
equal dimension may cancel almost completely in ${\cal R}_{\gamma ,\ell}$
and ${\cal R}_{\gamma ,\nu}$, if the couplings are similar in magnitude.
This is illustrated in Fig.~14, where we show ${\cal R}^{-1}_{\gamma ,\nu}$
versus $p_T^{\rm min}(\gamma)$ for the SM (solid line), and two
combinations of anomalous $WW\gamma$ and $ZZ\gamma$ couplings.
The error bars in Fig.~14 display the statistical
errors expected for $\int\!{\cal L}dt=25$~pb$^{-1}$ and $W\to e\nu$
decays. The dashed line shows the expected result for $\lambda_0=1.7$ and
$h^Z_{30}=1.5$. Both couplings correspond to operators of dimension~6 in
the effective Lagrangian. It is clear that,
for these couplings and with the integrated luminosity expected from
the current Tevatron run, the deviation from the SM cannot be seen. The
dotted line in Fig.~14 shows ${\cal R}^{-1}_{\gamma ,\nu}$ for
$\Delta\kappa_0=2.6$ and $h^Z_{40}=0.075$. $\Delta\kappa_0$ corresponds
to a dimension~4 operator, whereas $h^Z_{40}$ originates from an
operator of dimension~8 in the effective Lagrangian. At small
minimum photon transverse momenta, the effects of the anomalous $WW\gamma$
coupling dominate, and the inverse cross section ratio is
larger than expected in the SM. For larger values of $p_T^{\rm
min}(\gamma)$, the influence of the higher dimensional coupling on the
$Z\gamma$ cross section increases, and ${\cal R}^{-1}_{\gamma ,\nu}$ drops
below the SM value. Only for $p_T^{\rm min}(\gamma)\approx
100$~GeV do the effects of the two non-standard contributions cancel.
Although no substantial cancelations over an extended region of
$p_T^{\rm min}(\gamma)$ occur between $\Delta\kappa$ and $h^Z_4$, the
error bars in Fig.~14 indicate that
it will be difficult to discriminate between the SM prediction
and the dotted curve at a statistically significant level with the data
expected from the current Tevatron run.

Possible cancelations between anomalous $WW\gamma$ and
$ZZ\gamma/Z\gamma\gamma$ couplings in ${\cal R}_{\gamma ,\ell}$ and
${\cal R}_{\gamma ,\nu}$ can be excluded through a measurement of the ratios
${\cal R}_{W\gamma}$ and ${\cal R}_{Z\gamma}$. Since the three vector
boson vertices do not enter the quantity in the denominator, and
${\cal R}_{W\gamma}$ (${\cal R}_{Z\gamma}$) is only sensitive to
$WW\gamma$ ($ZZ\gamma /Z\gamma\gamma$) couplings, cancelations between
the effects of non-standard $WW\gamma$ and $ZZ\gamma /Z\gamma\gamma$
couplings cannot occur in these cross section ratios.
The SM result for ${\cal R}_{W\gamma}$ [${\cal R}_{Z\gamma}$] versus
$p_T^{\rm min}(\gamma)$ is compared to $\sigma(W\gamma)/\sigma(W)$
[$\sigma(Z\gamma)/\sigma(Z)$] in the presence of anomalous $WW\gamma$
[$ZZ\gamma$] couplings in Fig.~15a [Fig.~15b]. The error bars indicate
the statistical errors expected for 25~pb$^{-1}$, taking only the decays
$W\to e\nu$ and $Z\to e^+e^-$ into account. Due to the large
number of $W$ bosons expected, the statistical error of ${\cal
R}_{W\gamma}$ is considerably smaller than that of ${\cal
R}_{\gamma ,\ell}$ and ${\cal R}_{\gamma ,\nu}$. In the current Tevatron
run it should be possible to measure ${\cal R}_{W\gamma}$ for minimum
photon transverse
momenta of up to $p_T^{\rm min}(\gamma)\approx 50$~GeV. The sensitivity of
${\cal R}_{V\gamma}$ ($V=W,\, Z$) to anomalous couplings is quite
similar to that of ${\cal R}_{\gamma ,\nu}$. Similar to the situation
encountered for ${\cal R}_{\gamma ,\ell}$, deviations from the SM
predictions are less pronounced in ${\cal R}_{W\gamma}$
versus $m_{\rm min}$ than for the cross section ratio as a function
of $p_T^{\rm min}(\gamma)$.

\section{Sensitivity Limits}

As we have demonstrated so far, the cross section ratios listed in
Eqs.~(1.2) -- (1.5) are sensitive indicators of anomalous couplings. We
now want to make this statement more quantitative by deriving those
values of $\Delta\kappa_0$, $\lambda_0$, and $h^V_{i0}$ ($V=\gamma,\,Z$)
which would give rise to a deviation from the SM at the level of one or
two standard deviations in the various cross section ratios. We assume
an integrated luminosity of
25~pb$^{-1}$ at the Tevatron and the cuts listed in Eqs.~(2.4) -- (2.7).
For ${\cal R}_{\gamma ,\nu}$ the photon transverse momentum cut is
increased to $p_T(\gamma)>30$~GeV, in order to reduce backgrounds from
prompt photon and two jet production. Sensitivity limits are calculated
for form factors of the form given in Eq.~(3.9) with $\Lambda=750$~GeV,
$n=2$ for the $WW\gamma$ couplings $\Delta\kappa_0$ and $\lambda_0$, and
$n=3$ ($n=4$) for $h^V_{10,30}$ ($h^V_{20,40}$) ($V=\gamma,\, Z$).

Our analysis is based on cross section ratios obtained in the Born
approximation and takes into account the expected theoretical and
systematic uncertainties. Based on the results presented in Section~2.3, we
roughly estimate the combined theoretical and systematic uncertainties
from the parametrization of the
parton distribution functions, the choice of the factorization scale
$Q^2$, and higher order QCD corrections to be about 10\% for
${\cal R}_{\gamma ,\nu}$ and ${\cal R}_{\gamma ,\ell}$, 20\% for ${\cal
R}_{W\gamma}$, and approximately 15\% for ${\cal R}_{Z\gamma}$ for the
range of photon transverse momenta accessible in the current Tevatron
run. In order to obtain these numbers we have added the various
contributions in quadrature. Possible systematic errors originating from
background processes are ignored. In estimating the uncertainties from
higher order QCD corrections, we have assumed that the photon isolation
cut (2.13) and the central jet veto of Eq.~(2.14) are imposed in
addition to cuts of Eqs.~(2.4) -- (2.7).

{}From the discussion in Section~3.2 it is
clear that, in most cases, the best sensitivity limits are obtained if
the ratios are viewed as functions of the minimum photon transverse
momentum. In the following we therefore derive bounds only for cross
section ratios viewed as a function of $p_T^{\rm min}(\gamma)$.
In the ratios of $Z\gamma$ to $W^\pm\gamma$ cross sections we vary either
$WW\gamma$ or $ZZ\gamma$ couplings. However, interference effects between
$\Delta\kappa_0$ and $\lambda_0$, and between the various $ZZ\gamma$
couplings $h^Z_{i0}$, are fully taken into account in our analysis.
Interference effects between $ZZ\gamma$ and $Z\gamma\gamma$ couplings
are expected to be small\refmark{\BaBe} and are ignored.
Sensitivity limits for $h^\gamma_{i0}$ are nearly identical to those
derived for $h^Z_{i0}$. Furthermore, bounds for the $CP$ violating
couplings $h^Z_{10,20}$ virtually coincide with those for $h^Z_{30,40}$.
We therefore concentrate on $\Delta\kappa_0$, $\lambda_0$, $h^Z_{30}$,
and $h^Z_{40}$ in the following.

To estimate the sensitivity bounds which can be achieved at the
Tevatron, we use the maximum likelihood technique.
The likelihood function is calculated using binomial probability
distributions for the cross section ratios\rlap.\refmark{\JR} The minimum
photon transverse momentum is increased in steps of at least 5~GeV, starting at
$p_T^{\rm min}(\gamma)=10$~GeV for ${\cal R}_{V\gamma}$ and ${\cal
R}_{\gamma ,\ell}$, and at $p_T^{\rm min}(\gamma)=30$~GeV for ${\cal
R}_{\gamma ,\nu}$. For smaller steps in $p_T^{\rm min}(\gamma)$, the
cross section ratios for different minimum photon transverse momenta are
strongly correlated, resulting in overly optimistic sensitivity limits.

The resulting bounds for $\Delta\kappa_0$, $\lambda_0$, and $h^Z_{30,40}$ are
presented in Table~1. Due to the larger statistical errors in ${\cal
R}_{\gamma ,\ell}$, the limits achievable from this ratio are about 20
-- 30\% weaker than those from the other cross section ratios. The 95\%
CL bounds from ${\cal R}_{\gamma ,\nu}$ and ${\cal R}_{V\gamma}$ ($V=W,\,
Z$) are quite similar. The larger statistical errors in ${\cal
R}_{\gamma ,\nu}$ are almost completely compensated by the smaller
systematic and theoretical errors.
Table~1 clearly demonstrates the advantage of ${\cal R}_{\gamma ,\nu}$ due
to the larger branching ratio of the $Z\to\bar\nu\nu$ decay. The limits
on the $WW\gamma$ couplings $\Delta\kappa_0$ and $\lambda_0$ depend only
slightly on the form factor scale, whereas the bounds on $h^Z_{30,40}$
can easily change by a factor~3 --~6 if $\Lambda$ is varied by a
factor~2 (Ref.~\BaBe).

At Tevatron energies, nonnegligible interference effects are found
between $\Delta\kappa$ and $\lambda$, and $h^Z_3$ and $h^Z_4$. As a
result, different anomalous contributions to the helicity amplitudes may
cancel partially, resulting in weaker bounds than if only one coupling
at a time is allowed to deviate from its SM value. These
effects are fully taken into account in Table~1.
If only one coupling is varied at a time, the limits of Table~1 for
$\Delta\kappa_0$ and $\lambda_0$ improve by 10 --~30\%.
For example, one finds
$$ \Delta\kappa_0=0\matrix{+0.9\crc -0.7}~~({\rm for}~\lambda_0=0),~~{\rm
and}~~\lambda_0=0\matrix{+0.28\crc -0.29}~~({\rm for}~\kappa_0=1),
\eqno (3.11) $$
at the $1\sigma$ level from ${\cal R}_{\gamma ,\nu}$. With
$\int\!{\cal L}dt=25$~pb$^{-1}$, the present UA2 limit for $\kappa$
($\lambda$) [see Eq.~(3.5)] thus may be improved by up to a
factor~3~(5). For the form factor parameters used, the bounds for
$h^Z_{30}$ and $h^Z_{40}$ in Table~1 improve by a factor 1.6 --~2 if
only one coupling is varied at a time.

The sensitivity to anomalous couplings stems from regions of phase
space where the anomalous
contributions to the cross sections are considerably larger than the SM
expectation. As a result, the bounds scale essentially like
$\left(\int\!{\cal L}dt\right)^{1/4}$. Therefore, increasing the
integrated luminosity at the Tevatron to 100~pb$^{-1}$, as foreseen by
the end of 1994, will improve the sensitivity limits of Table~1 by about a
factor~1.4. Due to the smaller experimental, theoretical, and systematic
uncertainties of the cross section ratios, the resulting bounds may be
considerably better than those expected
from analyzing the $p_T(\gamma)$ distribution\rlap.\refmark{\BB,\BaBe}

The bounds listed in Table~1 have been obtained for a generic set of
cuts [Eqs.~(2.4) -- (2.7)]. They also depend somewhat on the exact
procedure used to extract the limits. For example, increasing
$p_T^{\rm min}(\gamma)$ in steps of 30~GeV, weakens the bounds by
20 -- 30\%.
Our limits thus should be regarded as guidelines, illustrating the
capabilities of CDF and D$\zero$ in improving our knowledge of
$WW\gamma$ and $ZZ\gamma /Z\gamma\gamma$ couplings within the immediate
future.

As we have mentioned before, for 25~pb$^{-1}$ the sensitivity of the cross
section ratios to anomalous couplings is limited mostly by statistical
errors. For this situation, a calculation of the ratios at tree level is
completely sufficient. For larger integrated luminosities, the
theoretical and systematic errors become more important in limiting the
sensitivity bounds which can be achieved. These errors
could be improved substantially if a full ${\cal O}(\alpha_s)$
calculation of the ratios for general $WW\gamma$ and $ZZ\gamma
/Z\gamma\gamma$ couplings is carried out. This would in particular
reduce the uncertainty originating from the choice of the
factorization scale $Q^2$, which dominates the systematic and
theoretical errors in ${\cal R}_{W\gamma}$ and
${\cal R}_{Z\gamma}$ at large $p_T^{\rm min}(\gamma)$.

\chapter{Summary and Conclusions}

In this paper we have studied the theoretical aspects of cross section
ratios for the processes $p\bar p\to W^\pm\gamma$ and $p\bar p\to
Z\gamma$ at Tevatron energies. Four different ratios can be formed,
which are listed in Eqs.~(1.2) -- (1.5). Compared to direct measurements
of cross sections, experimental, theoretical, and systematic errors are
expected to be significantly reduced in ratios of cross sections.

Our main results can be summarized as follows:

\item{1)} The ratios ${\cal R}_{\gamma ,\ell}=B(Z\to\ell^+\ell^-)\cdot
\sigma(Z\gamma)/\allowbreak B(W\to\ell\nu)\cdot\sigma(W^\pm\gamma)$ and
${\cal R}_{\gamma ,\nu}=B(Z\to\bar\nu\nu)\cdot\sigma(Z\gamma)/\allowbreak
B(W\to\ell\nu)\cdot\sigma(W^\pm\gamma)$ as a function of
the minimum photon transverse momentum, $p_T^{\rm min}(\gamma)$,
increase sharply with $p_T^{\rm min}(\gamma)$ in the SM,
reflecting the radiation zero which is present in the lowest order
$q\bar q'\to W^\pm\gamma$ helicity amplitudes.

\item{2)} The systematic and theoretical errors of ${\cal R}_{\gamma ,\ell}$
and ${\cal R}_{\gamma ,\nu}$ are significantly smaller than those of
${\cal R}_{V\gamma}=\sigma(V\gamma)/\sigma(V)$ ($V=W^\pm,Z$).
Theoretical and systematic uncertainties are well under control for all
cross section ratios.

\item{3)} Higher order QCD corrections only partially cancel in the
cross section ratios, in particular at large photon transverse momenta.
The imperfect cancelations can be traced to a phase space region where a
high $p_T$
photon is balanced by a quark jet which emits a $W$ or $Z$ boson almost
collinear with the quark. By applying a modest central jet veto
requirement [see Eq.~(2.14)], the residual QCD corrections cancel
almost completely in
the cross section ratios over a wide range of photon transverse
momenta.

\item{4)} The $W^\pm\gamma$ and $Z\gamma$ cross section ratios listed
in Eqs.~(1.2) -- (1.5) constitute powerful new tools which
can be used to set new limits on physics beyond the SM. We
have studied in detail the impact of non-standard $WW\gamma$ and
$ZZ\gamma/Z\gamma\gamma$ couplings on the cross section ratios and have
derived sensitivity limits (see Table~1) based on an integrated luminosity
of 25~pb$^{-1}$ expected from the current Tevatron run. For anomalous
$WW\gamma$ couplings, these limits improve present hadron collider
bounds up to a factor~3~--~5. The various cross section ratios yield
complementary information on the three vector boson~couplings.

The bounds listed in Table~1 should be compared with theoretical
expectations, existing low energy limits, and constraints obtained from
LEP~I data. In models based on chiral perturbation theory, for example,
one typically expects deviations from the SM of ${\cal
O}(10^{-2})$ (Ref.~\BDV). Although bounds can be extracted from
low energy and high precision measurements at the $Z$ pole, there are
ambiguities and model dependencies in the results\rlap.\refmark{\De -
\HISZ} From loop contributions
to $(g-2)_\mu$ one estimates\refmark{\muon} limits which are typically
of ${\cal O}(1-10)$. No rigorous bounds on $WW\gamma$ couplings can be
obtained from LEP~I data, if correlations between different
contributions to the anomalous couplings are fully taken into account.
Without serious cancelations among various one loop contributions, one
finds\refmark{\HISZ}
$$ |\Delta\kappa|,~|\lambda|\lsim 0.5-1.5 \eqno (4.1) $$
at the 90\% CL from present data on $S$, $T$, and $U$ (Ref.~\PT) [or,
equivalently,
$\epsilon_1$, $\epsilon_2$, and $\epsilon_3$ (Ref.~\Alt)]. The limits
which can be obtained from data expected in the current Tevatron
run are already competitive with the bounds of Eq.~(4.1). Constraints
on $ZZ\gamma$ and $Z\gamma\gamma$ couplings from $S$, $T$, and $U$ have
not been calculated so far. LEP~I data on radiative $Z$ decays provide
only very little information on the structure of the $ZZ\gamma
/Z\gamma\gamma$ vertex\rlap.\refmark{\BaBe}

Significant improvements of the bounds derived in Table~1 can be
expected
if an integrated luminosity of 100~pb$^{-1}$ is accumulated at the
Tevatron, as foreseen by the end of 1994, and from $W$ pair and
$Z\gamma$ production at LEP~II\rlap.\refmark{\HHPZ,\Foxl} Finally,
the LHC and SSC\rlap,\refmark{\BZ} and a linear $e^+e^-$ collider with
$\sqrt{s}=500$~GeV (Refs.~\Boud,\GG) will enable a
measurement of the $WW\gamma$ and $ZZ\gamma /Z\gamma\gamma$ couplings at
the 1\% level. In view of our present poor knowledge of the self
interactions of $W$ bosons, $Z$ bosons, and photons, the limits which can be
obtained from a measurement of the $W^\pm\gamma$ and $Z\gamma$ cross
section ratios with the data accumulated in the current Tevatron run will
represent a major step forward
towards a high precision measurement of the three vector boson vertices.

\vskip 0.35in
\ack
We would like to thank F.~Halzen, S.~Keller, G.~Landsberg, and
D.~Zeppenfeld for stimulating discussions, and encouragement. We are
also grateful to H.~Baer for providing us with the
FORTRAN code of Ref.~\BR. One of us (UB) would like to thank the
Fermilab Theory
Group, where this work was completed, for its warm hospitality.
This research was supported in part by the UK
Science and Engineering Research Council, and in part by the
U.~S.~Department of Energy under Grant No.~DE-FG02-91ER40677 and
Contract No.~DE-FG05-87ER40319.
\endpage
\par \penalty-400 \vskip\chapterskip
   \spacecheck\referenceminspace \immediate\closeout\referencewrite
   \referenceopenfalse
   \line{\fourteenrm\hfil REFERENCES\hfil}\vskip\headskip
   \input referenc.texauxil
   
\endpage
\centerline{TABLE~1}
\vskip 3.mm
\noindent
Sensitivities achievable at the $1\sigma$ and $2\sigma$ confidence levels
(CL) for the anomalous $WW\gamma$ and $ZZ\gamma$ couplings
$\Delta\kappa_0$, $\lambda_0$, $h_{30}^Z$, and $h_{40}^Z$ from the cross
section ratios ${\cal R}_{\gamma ,\ell}$, ${\cal R}_{\gamma ,\nu}$,
${\cal R}_{W\gamma}$, and ${\cal R}_{Z\gamma}$, for an
integrated luminosity of 25~pb$^{-1}$ at the Tevatron.
The procedure used to extract the sensitivity bounds is
described in the text. The limits
for $\Delta\kappa_0$ ($h_{30}^Z$) apply for arbitrary values of
$\lambda_0$ ($h_{40}^Z$) and vice versa. For the form
factors we use Eq.~(3.9) with $\Lambda=750$~GeV, $n=2$ for $WW\gamma$
couplings, and $n=3$ ($n=4$) for $h_{30}^Z$ ($h_{40}^Z$), respectively.
The $W$ and $Z$ decay channels into muons are not included in deriving the
sensitivity limits. Anomalous $Z\gamma\gamma$ couplings are assumed to
be zero.
\vskip 0.5in
\tablewidth=5.5in
\def\tstrut{\vrule height 4.3ex depth 2.7ex width 0pt}
\begintable
coupling | CL | ${\cal R}_{\gamma ,\ell}$ | ${\cal R}_{\gamma ,\nu}$
| ${\cal R}_{W\gamma}$ \crthick
$\Delta\kappa_0$ | $2\sigma$ | $\matrix{+2.5 \crc -2.0}$ |
$\matrix{+1.7 \crc -1.3}$ | $\matrix{+1.7 \crc -1.3}$ \nr
 | $1\sigma$ | $\matrix{+1.8 \crc -1.3}$ | $\matrix{+1.2 \crc -0.9}$ |
$\matrix{+1.5 \crc -1.1}$ \cr
$\lambda_0$ |$2\sigma$ | $\matrix{+0.84 \crc -0.98}$ |
$\matrix{+0.49 \crc -0.57}$ | $\matrix{+0.52 \crc -0.60}$ \nr
 | $1\sigma$ | $\matrix{+0.54 \crc -0.69}$ | $\matrix{+0.32 \crc -0.40}$
 | $\matrix{+0.44 \crc -0.55}$  \crthick
coupling | CL | ${\cal R}_{\gamma ,\ell}$ | ${\cal R}_{\gamma
,\nu}$ | ${\cal R}_{Z\gamma}$ \crthick
$h_{30}^Z$  | $2\sigma$ |$\pm 1.0$ | $\pm 0.8$ | $\pm 0.9$ \nr
 | $1\sigma$ |$\pm 0.7$ | $\pm 0.5$ | $\pm 0.7$ \cr
$h_{40}^Z$  | $2\sigma$ | $\pm 0.16$ | $\pm 0.13$ | $\pm 0.14$ \nr
 | $1\sigma$ | $\pm 0.11$ | $\pm 0.09$ | $\pm 0.11$
\endtable
\endpage
\par \penalty-400 \vskip\chapterskip
   \spacecheck\referenceminspace \immediate\closeout\figurewrite
   \figureopenfalse
   \line{\fourteenrm\hfil FIGURE CAPTIONS\hfil}\vskip\headskip
   \input figures.texauxil
   
\endpage
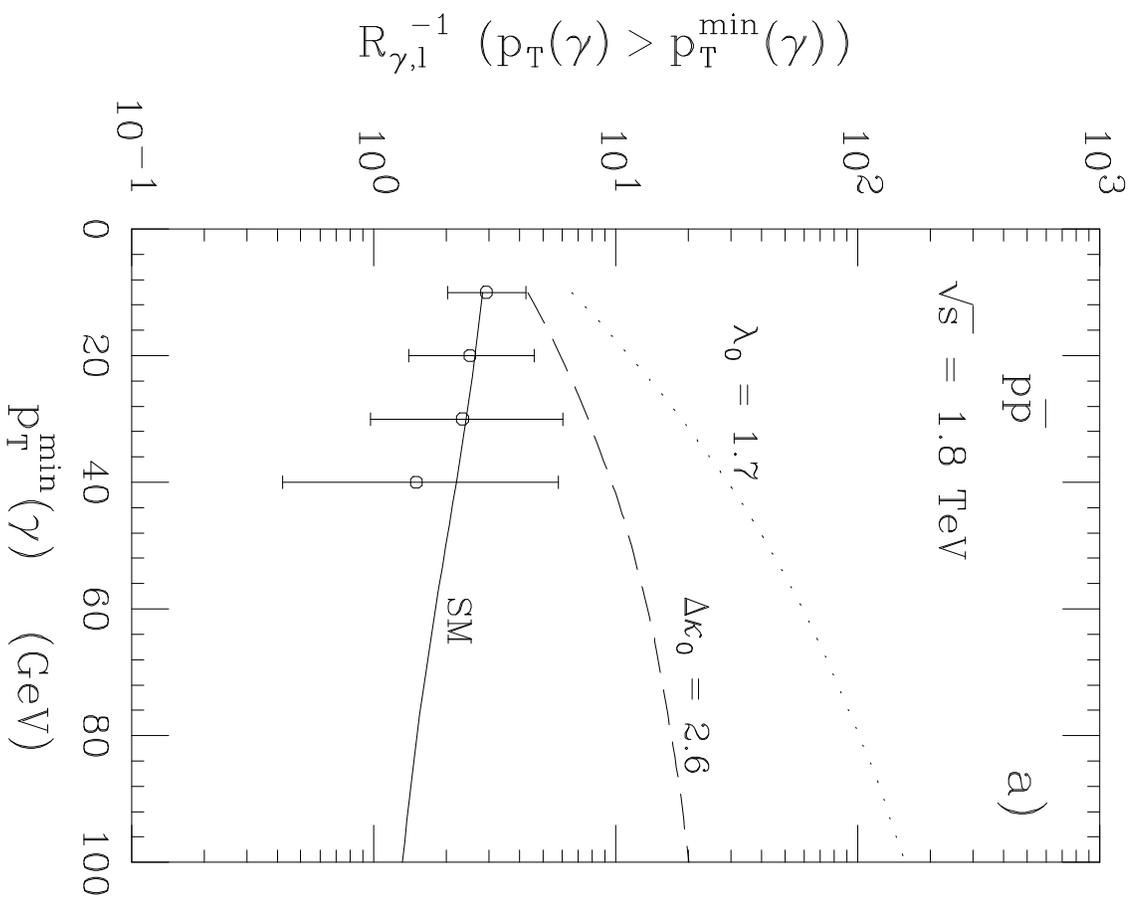
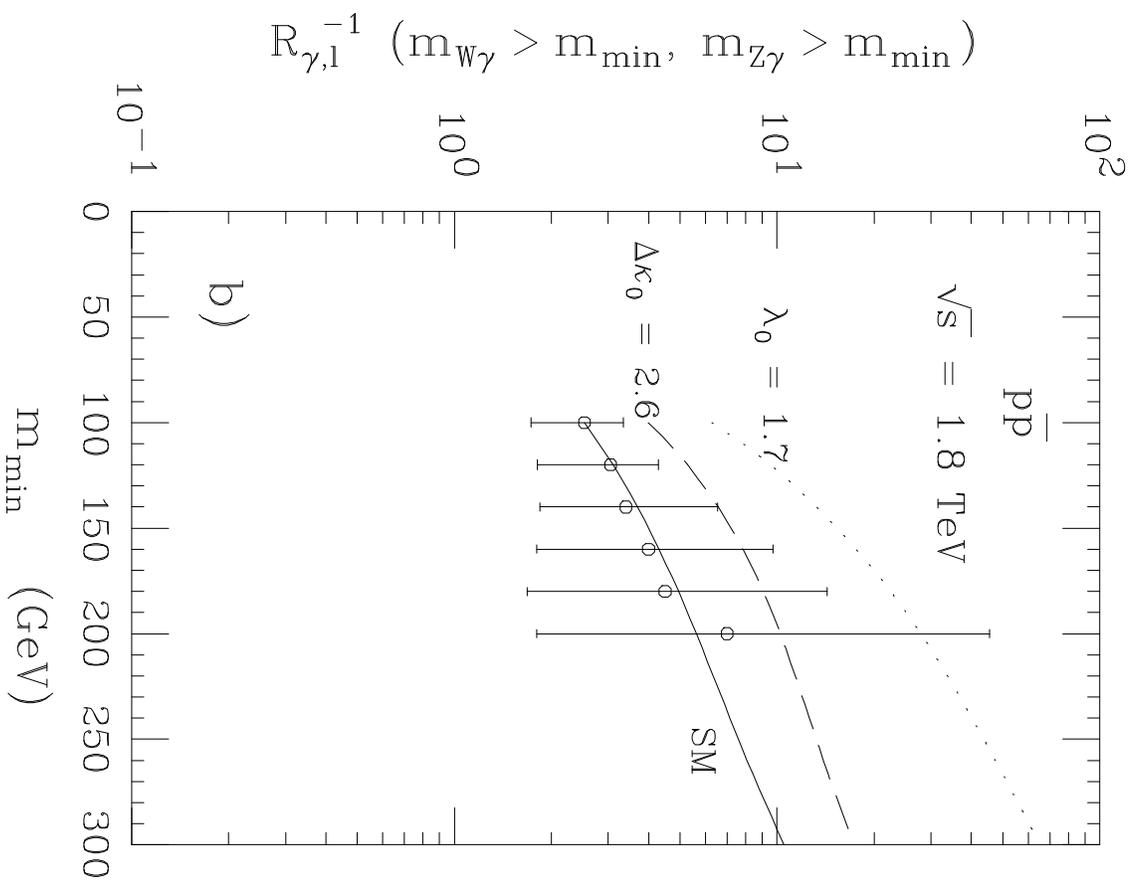
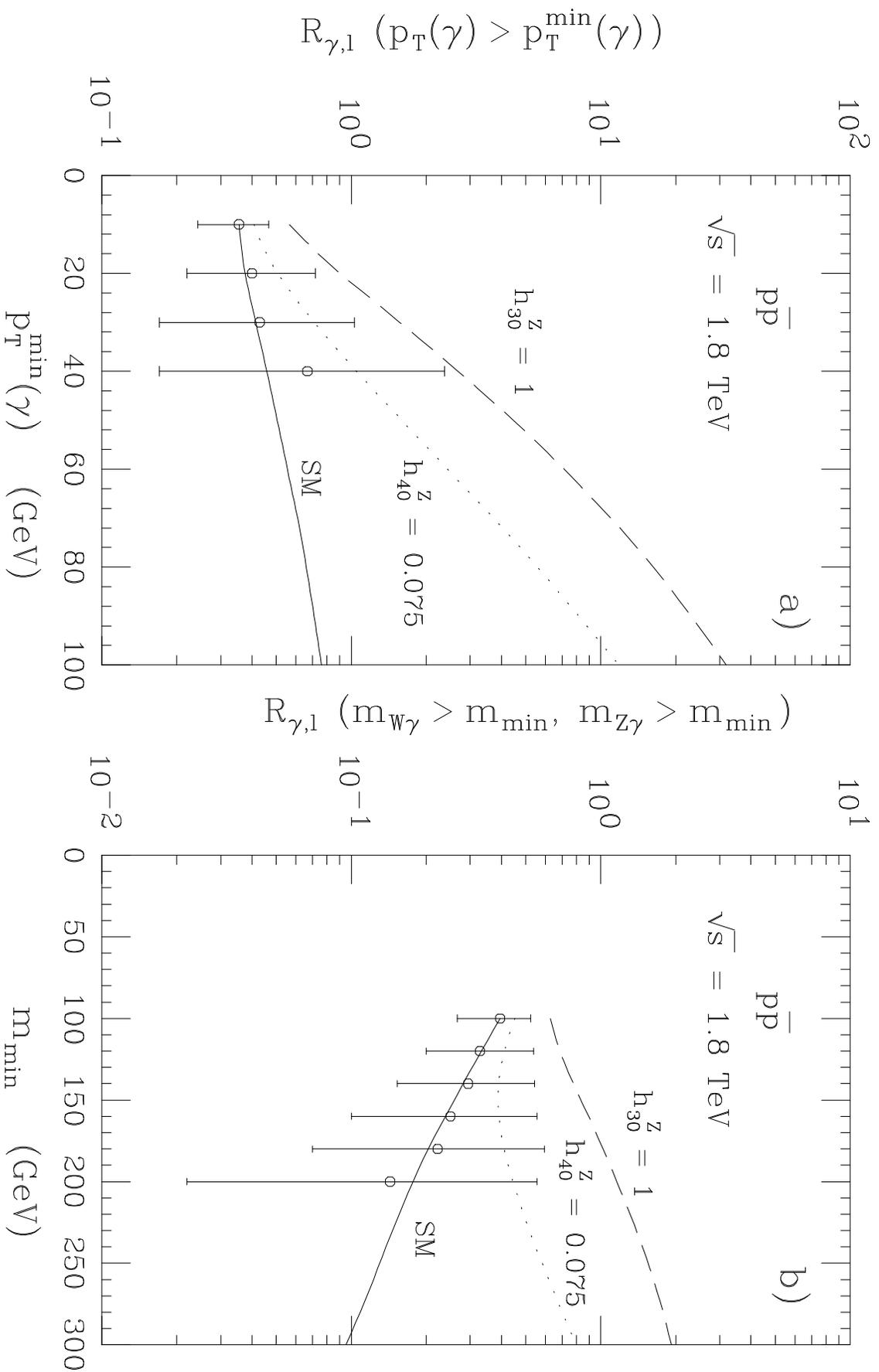
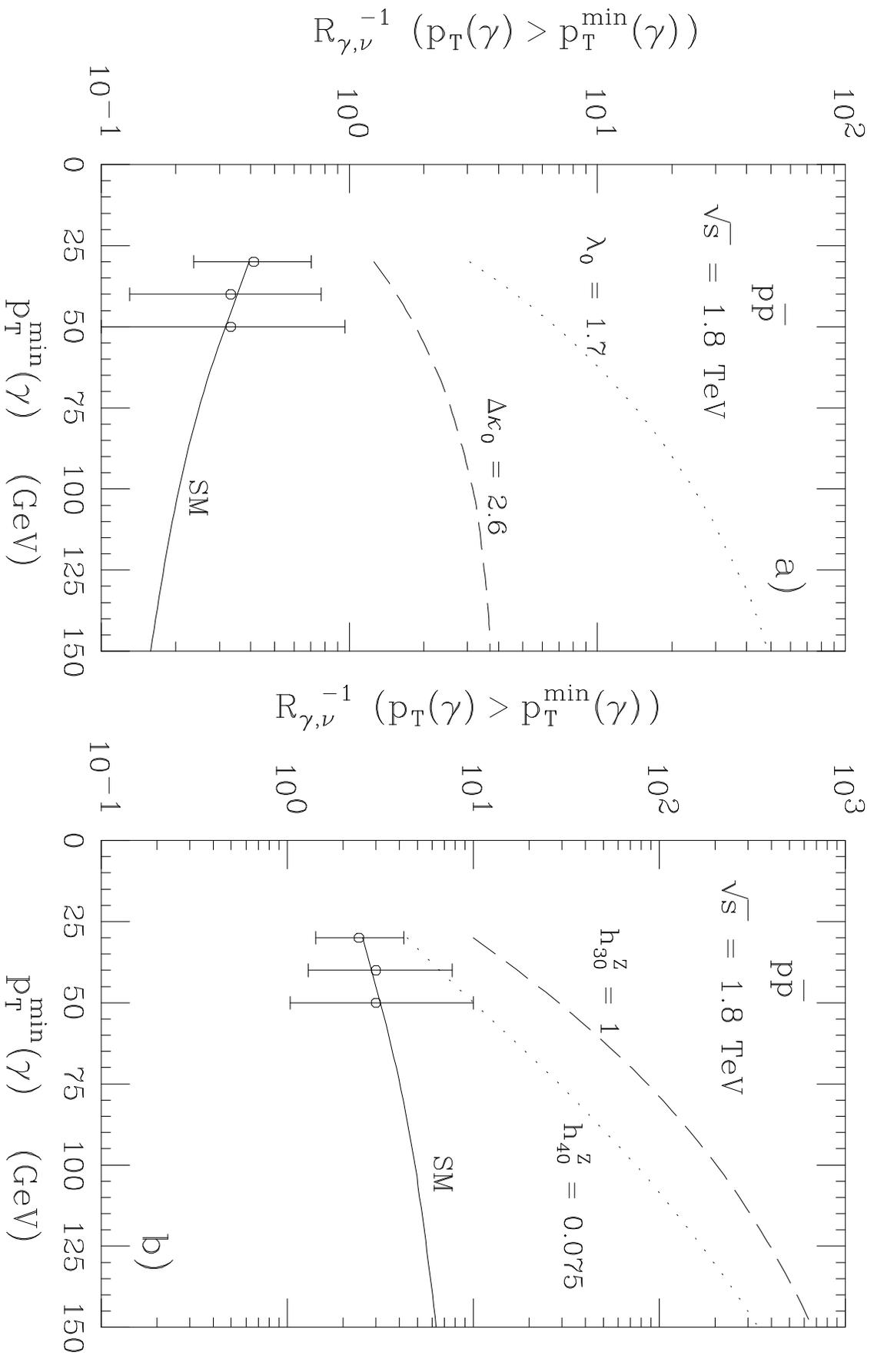
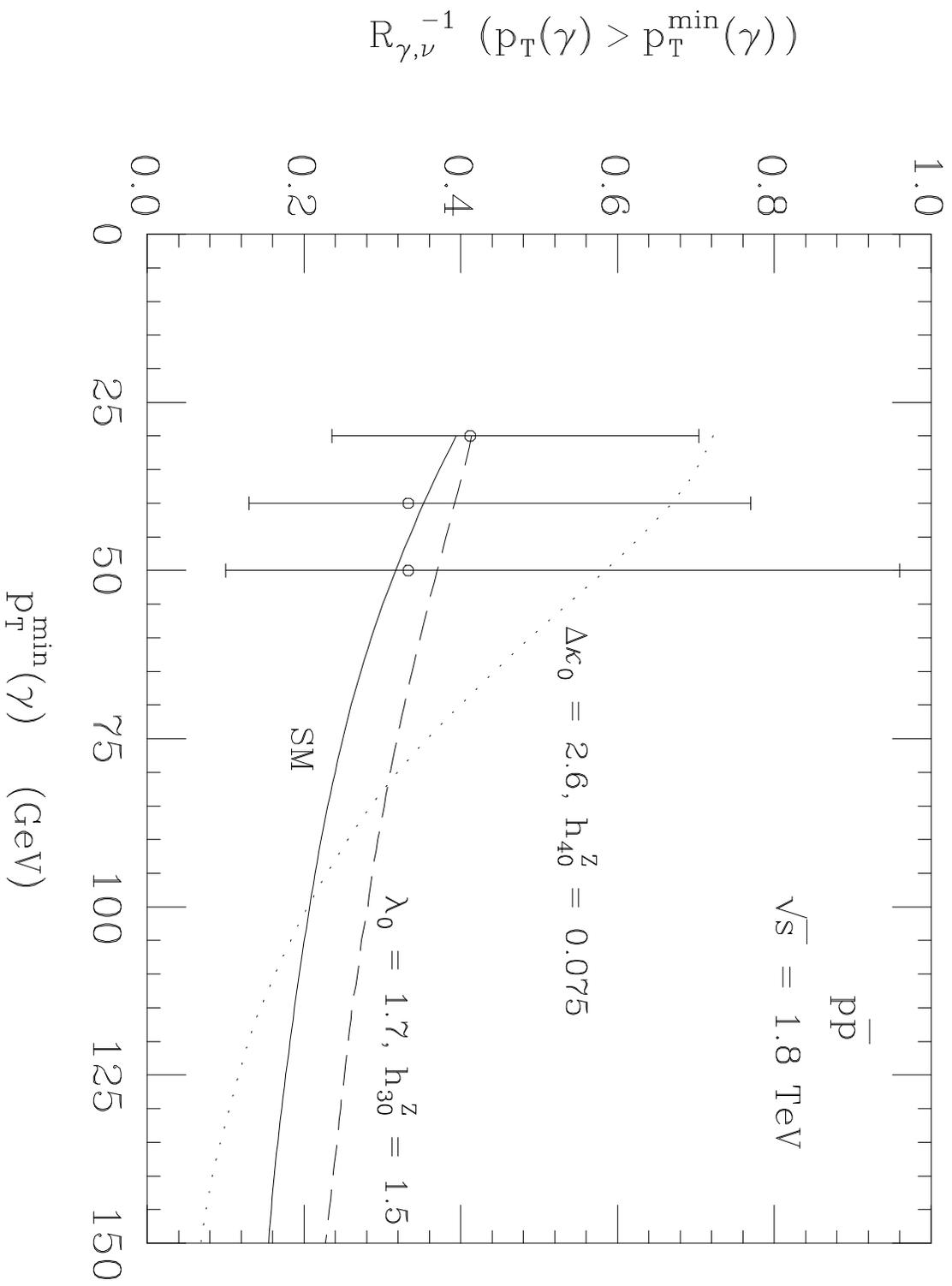
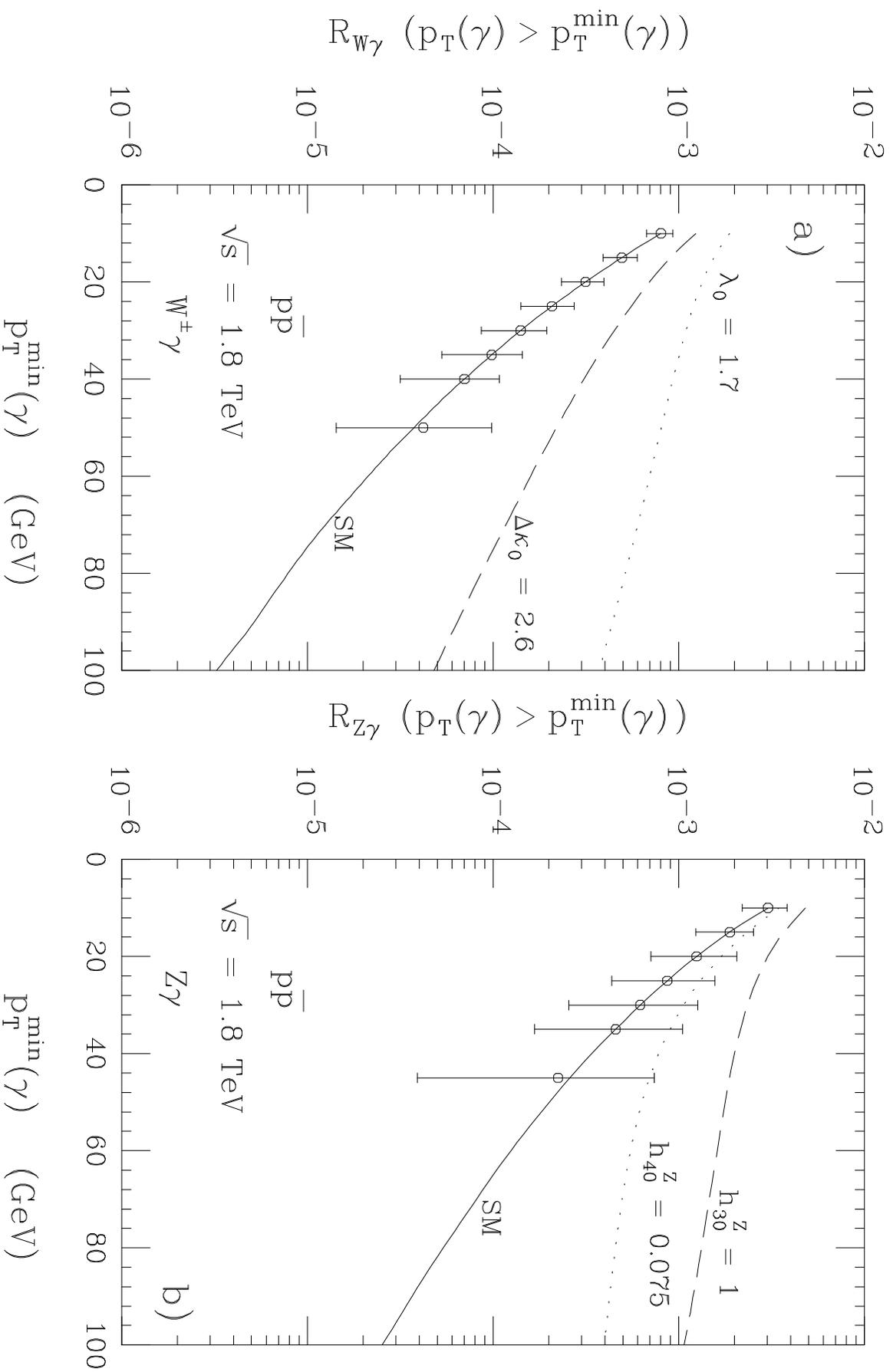
\bye